\documentclass[aps,pra,twocolumn,showpacs]{revtex4}
\usepackage{amssymb}
\usepackage{graphicx}
\usepackage{amsmath}
\usepackage{dcolumn}
\usepackage{bm}

\newcommand{\beq}{\begin{equation}}
\newcommand{\eeq}{\end{equation}}
\newcommand{\beqa}{\begin{eqnarray}}
\newcommand{\eeqa}{\end{eqnarray}}

\def\ajp#1{{ Am.\ J.\ Phys.} {\bf #1}}
\def\epl#1{{ Europhys.\ Lett.} {\bf#1}}

\def\oc#1{{ Opt.\ Commun.} {\bf#1}}

\def\jpb#1{{ J.\ Phys.\ B} {\bf#1}}
\def\jpa#1{{ J.\ Phys.\ A} {\bf#1}}

\def\pla#1{{ Phys.\ Lett. A\/} {\bf#1}}

\def\pra#1{{ Phys.\ Rev. A\/} {\bf#1}}

\def\prl#1{{ Phys.\ Rev.\ Lett.} {\bf#1}}
\def\qic#1{{ Quant.\ Inf.\ Comp.} {\bf#1}}
\def\sci#1{{ Science} {\bf#1}}
\def\rmp#1{{ Rev. \ Mod. \ Phys.} {\bf#1}}

\begin{document}

\title{Pre-management of Disentanglement}
\author{Xiao-Feng Qian}
\email{xfqian@pas.rochester.edu}
\affiliation{Rochester Theory Center and Department of Physics \& Astronomy\\
University of Rochester, Rochester, New York 14627}
\date{\today }

\begin{abstract}
We define threshold disentanglement (TD) as the level above which entanglement must be maintained for later use and calculate the values of quantum system parameters at $t=0$ that will ensure that the threshold is not crossed in a specified time. An important issue is the possibility of non-smooth, typically non-exponential, decay of entanglement into the state of complete disentanglement (CD). Given a specified operation time $\tau$, we determine phase diagrams showing entanglement preservation during the evolution of a generic two-qubit mixed state as a function of selected initial parameters such as concurrence and excitation probabilities. In other words, disentanglement can be pre-managed through these parameters. Three common relaxation processes - amplitude, phase and depolarization - are considered.
\end{abstract}

\pacs{03.65.Ud, 03.65.Yz, 42.50.Pq}
\maketitle

\section{Introduction}
Preservation of memory is an important goal for every kind of coherent quantum process. The type of control that this implies becomes more difficult in cases involving multi-particle cooperation, typically including entanglement in some way. Preservation of entanglement is often an important goal itself. Since dissipation processes induced by natural environments will eventually or quickly take entanglement all the way to zero, two related criteria emerge for consideration. These are (a) the minimum degree of entanglement that is still ``useful", setting a lower bound for entanglement, a threshold disentanglement (TD) value that must not be crossed, and (b) the length of time during which this useful level must be maintained.  For example, in order for a quantum computation task to be successful, the decay time to reach the low entanglement threshold must be longer than the time scale for computation.

Natural environments induce well-known single-party decoherence times ($T_1,\ T_2$, etc.), but these are often not helpful or even highly misleading because it is well known that disentanglement can, and often does, proceed by its own distinct evolution channel and is notoriously able, in only a finite time, to reach complete disentanglement (CD), colloquially known as entanglement sudden death \cite{Yu-Eberly09, Almeida-etal07}. Clearly a CD event may possibly violates the TD value at a given time. This has led a number of entanglement restoration methods to be proposed, using schemes of local unitary operations \cite{Rau-etal08}, quantum error correction \cite{ESD-QEC}, feedback control \cite{Li-etal11}, dynamical decoupling \cite{Goyal-etal12}, and environment assisted error correction \cite{Yu-etal}, in order to control and restore entanglement during environmental exposure. The proposals \cite{ESD-QEC, Li-etal11, Goyal-etal12, Rau-etal08} have shown that while these schemes may delay or even avert the occurrence of CD in some cases, they may also counter-intuitively accelerate or even induce CD. The proposal by Zhao, et al.~\cite{Yu-etal} relies on finding random unitary or non-random unitary Kraus decompositions, which are relatively difficult to achieve and are absent for universal decay channels. Proposals of entanglement preservation based on dynamical single-qubit population trapping have also been presented via the exploitation of structured quantum environments or suitable classical environments \cite{preservation} and experimental confirmation has also been realized \cite{preservation-exp}.

A second category of strategies is disentanglement prevention, in which specific setups are made before the disentanglement process. One typical approach of this category is to establish a decoherence free subspace \cite{Palma96, Zanardi-Rasetti97, Lidar98}, which requires either increasing the number of non-computing redundant qubits to create certain symmetry, or some engineering of the system or reservoir with serious external contacts. Another approach is to control disentanglement through initial conditions, i.e., during the state preparation process. Isolated connections between CD time and initial conditions have been identified extensively (see for example \cite{Yu-Eberly04, Ann-Jaeger07, Lopez08, Novotny11, Julian-etal12, Franco-etal13, Yu-Eberly07, Bellomo08, Man-etal08, Cui-etal09, Pang-etal12, Duan-etal13} and a review \cite{Yu-Eberly09}). Huang and Zhu derived the necessary and sufficient conditions for CD for a generic two-qubit state under amplitude damping and phase damping channels through the positivity requirement of seven principal minors of a $4\times4$ matrix \cite{Huang-Zhu07}. Fanchini et al.~\cite{Fanchini-etal11} have extended the results of Huang and Zhu to more general dissipative Hamiltonians. However, such CD conditions engage almost every matrix element of a given two-qubit initial state through various matrix transformations, and are extremely complex to control in reality.

Recently, we have shown that the susceptibility to CD can be controlled via initial entanglement and purity for a simple type of two-qubit states that undergo symmetric amplitude damping dissipation \cite{Qian-Eberly12}. This opens the possibility of pre-managing disentanglement properties with just a few controllable initial resources.  Yang et al.~\cite{Yang-etal13} followed this by showing that the CD pre-management procedure is also viable for extended Werner-like initial states. However, for practical situations, initial states may be arbitrarily mixed and decay channels can be rather versatile and complex. An open challenge arises: is it still possible to manage CD with only limited controllable initial parameters for practical complex situations. A further question follows naturally: is it also possible to include the threshold disentanglement condition within the control process?

An important complication is that even for the simple case of X-form matrices (first systematically considered by Yu and Eberly \cite{Yu-Eberly07}), the CD time can only be found for specific types of states, such as Werner-like states, ansatz states, etc., or symmetric reservoir interactions (see for example \cite{Yu-Eberly04, Yu-Eberly07, Bellomo08, Man-etal08, Cui-etal09, Pang-etal12, Duan-etal13}). The complication is mainly due to the fact that the disentanglement process, as well as measures of entanglement such as Schmidt weight \cite{Schmidt}, concurrence \cite{Wootters}, negativity \cite{negativity}, etc., are usually non-linear functions of time for given initial conditions.  This is because these non-linear dependences usually correspond to higher order linear equations that are difficult to solve analytically (see our detailed analysis in section III). For example, no analytic CD time is obtained for arbitrary X-form states with non-symmetric amplitude damping and depolarization channels.

However, here we make use of the fact that such difficulties can be avoided if one considers the inverse of the problem. That is, we will obtain the state's initial conditions for an arbitrarily given CD or TD time. This corresponds to practical situations when a particular quantum information task requires CD or TD to happen no earlier than a prescribed time $\tau$. Thus, in this paper we extend our previous analysis of CD pre-management \cite{Qian-Eberly12} by considering arbitrary two-qubit mixed states, and by treating non-symmetric damping of the two qubits for three different quantum channels, i.e., amplitude damping, phase damping, and depolarization. At the same time, we show explicitly for the first time the pre-management of threshold disentanglement properties and illustrate their connections with CD.

We first analyze the entanglement dynamics of an arbitrary X-form two-qubit mixed state. Exact CD-free, CD-no-go, CD-tolerable, TD-no-go, and TD-tolerable phases are explicitly determined in terms of several initial parameters, i.e., concurrence and excitation
probabilities. Two types of intuitive and counter-intuitive situations are discussed. We then extend the static entanglement lower bound relation for X-form matrices, developed recently by Ma et al.~\cite{Ma-etal11} and Hashemi
Rafsanjani et al.~\cite{Rafsanjani-Agarwal12}, to the dynamical
evolution of the state. We show that CD and TD phase critical conditions of
a generic state are lower bounded by these of its X-form component. Therefore the control of different CD and TD properties of a general
complex mixed state can be realized sufficiently through just a few initial parameters of its X-form component.

The paper is organized as follows: In section II, we demonstrate the time
evolution of a generic two qubit state under three quantum channels.
The dynamics of X-form mixed states are shown explicitly. Section III gives a
detailed demonstration of CD and TD phases for X-form states in all three
quantum channels. In section IV, the static entanglement lower bound
conditions between the X-form and the entire matrix is generalized to the
dynamical process, and is then used to determine the lower
bounds of CD and TD phases for the generic time dependent state. It is then
followed by summary and acknowledgement sections.

\section{Initial State and Time Evolution}

In spite of the rapid development of multi-partite and multi-dimensional
entanglement studies recently (see for example \cite{Kimble-etal09,
Davidovich-etal12, Huber13}, and references therein), qubit-qubit
entanglement remains as the central element for most quantum information and
quantum computation technologies \cite{NC-Preskill, Horodecki09-etal}. In practice, real
state preparations may be inevitably deviated from ideal target states.
Therefore, we consider the most general form of a prepared two-qubit ($A$
and $B$) mixed initial state, i.e.,
\begin{equation}
\rho (0)=\left(
\begin{array}{cccc}
\rho _{11} & \rho _{12} & \rho _{13} & \rho _{14} \\
\rho _{21} & \rho _{22} & \rho _{23} & \rho _{24} \\
\rho _{31} & \rho _{32} & \rho _{33} & \rho _{34} \\
\rho _{41} & \rho _{42} & \rho _{43} & \rho _{44}
\end{array}
\right),  \label{general initial}
\end{equation}
which is expressed in the standard basis: $|e\rangle _{A}|e\rangle _{B}$, $|e\rangle _{A}|g\rangle _{B}$, $|g\rangle _{A}|e\rangle _{B}$, $|g\rangle
_{A}|g\rangle _{B}$. Here $|e\rangle $ and $|g\rangle $ are usually
interpreted as excited and ground states of qubits $A$, $B$. It is obvious
that such a matrix can always be decomposed into a sum of X-form and O-form
matrices, i.e., $\rho (0)=X(0)+O(0)$. Here
\begin{equation}
X(0)=\left(
\begin{array}{cccc}
\rho _{11} & 0 & 0 & \rho _{14} \\
0 & \rho _{22} & \rho _{23} & 0 \\
0 & \rho _{32} & \rho _{33} & 0 \\
\rho _{41} & 0 & 0 & \rho _{44}
\end{array}
\right),  \label{X-form initial}
\end{equation}
and its non-zero matrix elements form an ``X" shape, and
\begin{equation}
O(0)=\left(
\begin{array}{cccc}
0 & \rho _{12} & \rho _{13} & 0 \\
\rho _{21} & 0 & 0 & \rho _{24} \\
\rho _{31} & 0 & 0 & \rho _{34} \\
0 & \rho _{42} & \rho _{43} & 0
\end{array}
\right)  \label{O-form initial}
\end{equation}
contains the remaining non-zero matrix elements that form an ``O" shape.

Another important practical aspect is the decoherence effect
caused by unavoidable environmental interactions after state preparation.
For a relatively complete illustration, we study three most commonly
considered quantum channels, i.e., amplitude damping, phase damping and
depolarization, that represent various physical system-reservoir interactions.

We consider the case when each qubit is independently interacting with a
local reservoir. Then the evolution of the two-qubit initial state can be
straightforwardly described by Kraus operator representations \cite{NC-Preskill}, i.e.,
\begin{equation}
\rho (t)=\sum_{i=1}K_{i}(t)\rho (0)K_{i}^{\dag }(t),  \label{generic time Kraus}
\end{equation}
where the joint Kraus operators $K_{i}(t)$, satisfying the relation $\sum_{i}K_{i}(t)K_{i}^{\dag }(t)=1$, correspond to all possible tensor
products of individual Kraus operators of qubits $A$, $B$, i.e., $K_{i}(t)=K_{m}^{A}\otimes K_{n }^{B}$. Here $K_{m}^{A}$ and $K_{n}^{B}$
represent all possible Kraus operators for qubits $A$ and $B$ respectively.

In the following we will demonstrate explicitly for all
three quantum channels the evolution of X-form state (\ref{X-form initial}), which is a valid density matrix by itself. So we denote the initial state with $\rho_{X}(0)$, and the time dependent state is obtained as $\rho_{X}(t)=\sum_{i=1}K_{i}(t)\rho_{X}(0)K_{i}^{\dag }(t)$.

\subsection{Amplitude Damping}

An amplitude damping channel can schematically model physical processes such
as spontaneous emission, spin chain Heisenberg interactions, etc. It
describes a situation of energy dissipation from the excited state of a two
level system to its ground state. The corresponding Kraus operators for such
a damping qubit, e.g., qubit $A$, are given as
\begin{equation}
K_{0}^{A}=\left(
\begin{array}{cc}
\sqrt{q_{a}} & 0 \\
0 & 1
\end{array}
\right) ,\quad K_{1}^{A}=\left(
\begin{array}{cc}
0 & 0 \\
\sqrt{p_{a}} & 0
\end{array}
\right),  \label{Kraus-Amplitude}
\end{equation}
where $p_{a}$ is the excitation-transfer probability that grows from $0$ to $1$ irreversibly, and $q_{a}=1-p_{a}$ denotes the probability that qubit $A$ remains in its excited state. The Kraus operators for qubit $B$ are
characterized similarly with a decay probability $q_{b}$.

By applying the Kraus operators (\ref{Kraus-Amplitude}) to the initial state
$\rho_{X}(0)$ one obtains the explicit time evolution as
\begin{equation}
\rho _{X}(t)=\left(
\begin{array}{cccc}
\rho _{11}(t) & 0 & 0 & \rho _{14}(t) \\
0 & \rho _{22}(t) & \rho _{23}(t) & 0 \\
0 & \rho _{32}(t) & \rho _{33}(t) & 0 \\
\rho _{41}(t) & 0 & 0 & \rho _{44}(t)
\end{array}
\right),  \label{X-amplitude time}
\end{equation}
where $\rho _{11}(t)=\rho _{11}q_{a}q_{b}$, $\rho _{22}(t)=\rho
_{11}q_{a}p_{b}+\rho _{22}q_{a}$, $\rho _{33}(t)=\rho _{11}p_{a}q_{b}+\rho
_{33}q_{b}$, $\rho _{44}(t)=\rho _{11}p_{a}p_{b}+\rho _{22}p_{a}+\rho
_{33}p_{b}+\rho _{44}$, $\rho _{14}(t)=\rho _{41}^{\ast }(t)=\rho _{14}\sqrt{q_{a}q_{b}}$, and $\rho _{23}(t)=\rho _{32}^{\ast }(t)=\rho _{23}\sqrt{q_{a}q_{b}}$. One notes that as time goes to infinity, the
population will be transferred completely into the ground state, i.e., the
matrix element $\rho _{44}(t)$.

\subsection{Phase Damping}

A phase damping channel characterizes noisy processes where relative phase
coherence in a system decays over time due to perturbation. It models
various physical situations such as photon scattering in a waveguide,
electronic state perturbations in an atom, etc. The Kraus operators for a
phase damping qubit are given as, say for qubit $A$,
\begin{equation}
K_{0}^{A}=\left(
\begin{array}{cc}
1 & 0 \\
0 & \sqrt{q_{a}}
\end{array}
\right), K_{1}^{A}=\left(
\begin{array}{cc}
0 & 0 \\
0 & \sqrt{p_{a}}
\end{array}
\right),  \label{Kraus-Phase}
\end{equation}
where $p_{a}$ is the probability that the system has been scattered off by
the reservoir, $q_{a}$ is the probability remain unchanged. The Kraus operators
of qubit $B$ has a scattering probability $p_{b}$.

Given the Kraus operators, the time evolution of the initial state $\rho_{X}(0)$ under phase damping can be easily obtained as
\begin{equation}
\rho _{X}(t)=\left(
\begin{array}{cccc}
\rho _{11} & 0 & 0 & \rho _{14}(t) \\
0 & \rho _{22} & \rho _{23}(t) & 0 \\
0 & \rho _{32}(t) & \rho _{33} & 0 \\
\rho _{41}(t) & 0 & 0 & \rho _{44}
\end{array}
\right),  \label{X-phase time}
\end{equation}
where $\rho _{14}(t)=\rho _{41}^{\ast }(t)=\sqrt{q_{a}q_{b}}\rho _{14}$, and
$\rho _{23}(t)=\rho _{32}^{\ast }(t)=\sqrt{q_{a}q_{b}}\rho _{23}$. Since the
phase damping perturbation only causes off-diagonal coherence loss, the
diagonal terms remain fixed in the time dependent state.

\subsection{Depolarization}

Depolarization is another important type of decoherence channel that
describes physical situations when a system (qubit) suffers bit flip and
phase flip errors from environmental interactions. It is a process in which
a qubit density matrix remains intact with probability $1-p$, and changes
into a completely unpolarized (depolarized) state $\hat{1}/2$ with
probability $p\in \lbrack 0,1]$. Here $\hat{1}$ represents a unit matrix.
For a depolarizing qubit $A$, there are four Kraus operators,
\begin{eqnarray}
K_{0}^{A} &=&\sqrt{q_{a}}\hat{1},K_{1}^{A}=\sqrt{p_{a}/3}\sigma _{1},  \notag
\\
K_{2}^{A} &=&\sqrt{p_{a}/3}\sigma _{2},K_{3}^{A}=\sqrt{p_{a}/3}\sigma _{3},
\label{Kraus-polarization}
\end{eqnarray}
where $p_{a}=3p/4$ is a re-scaled probability, $q_{a}=1-p_{a}$, and $\sigma _{i}$ with $i=1,2,3$ are the usual Pauli
matrices. The Kraus operators of qubit $B$ has a re-scaled depolarizing
probability $p_{b}$.

One then can straightforwardly achieve the time dependence of the X-form
initial state $\rho_{X}(0)$ under depolarization channel, i.e.,
\begin{equation}
\rho _{X}(t)=\left(
\begin{array}{cccc}
\rho _{11}(t) & 0 & 0 & \rho _{14}(t) \\
0 & \rho _{22}(t) & \rho _{23}(t) & 0 \\
0 & \rho _{32}(t) & \rho _{33}(t) & 0 \\
\rho _{41}(t) & 0 & 0 & \rho _{44}(t)
\end{array}
\right),  \label{X-polarization time}
\end{equation}
where $\rho _{14}(t)=\rho _{41}^{\ast }(t)=\rho _{14}f_{0}(t)$, $\rho
_{23}(t)=\rho _{32}^{\ast }(t)=\rho _{23}f_{0}(t)$, and
\begin{equation}
\left(
\begin{array}{c}
\rho _{11}(t) \\
\rho _{22}(t) \\
\rho _{33}(t) \\
\rho _{44}(t)
\end{array}
\right) =\left(
\begin{array}{cccc}
f_{1}(t) & f_{2}(t) & f_{3}(t) & f_{4}(t) \\
f_{2}(t) & f_{1}(t) & f_{4}(t) & f_{3}(t) \\
f_{3}(t) & f_{4}(t) & f_{1}(t) & f_{2}(t) \\
f_{4}(t) & f_{3}(t) & f_{2}(t) & f_{1}(t)
\end{array}
\right) \left(
\begin{array}{c}
\rho _{11} \\
\rho _{22} \\
\rho _{33} \\
\rho _{44}
\end{array}
\right).
\end{equation}
Here we have set $f_{0}(t)=q_{a}q_{b}-q_{a}p_{b}/3-p_{a}q_{b}/3+p_{a}p_{b}/9$, $f_{1}(t)=(1+2q_{a}+2q_{b}+4q_{a}q_{b})/9$, $f_{2}(t)=(2p_{b}+4q_{a}p_{b})/9$, $f_{3}(t)=(2p_{a}+4p_{a}q_{b})$, and $f_{4}(t)=4p_{a}p_{b}/9$. We see that the time
dependent matrix $\rho _{X}(t)$ will evolve into a completely unpolarized
state $\hat{1}/2$ as time goes to infinity.

We have shown that for all three channels the time evolution of the initial state $\rho_{X}(0)$ retains an
``X" form \cite{Yu-Eberly04, Yu-Eberly07, Bellomo08, Man-etal08, Cui-etal09, Pang-etal12, Duan-etal13}. This property plays an important role for the analysis of CD and TD phases in following sections.

\begin{figure}[t]
\includegraphics[width=4.2cm]{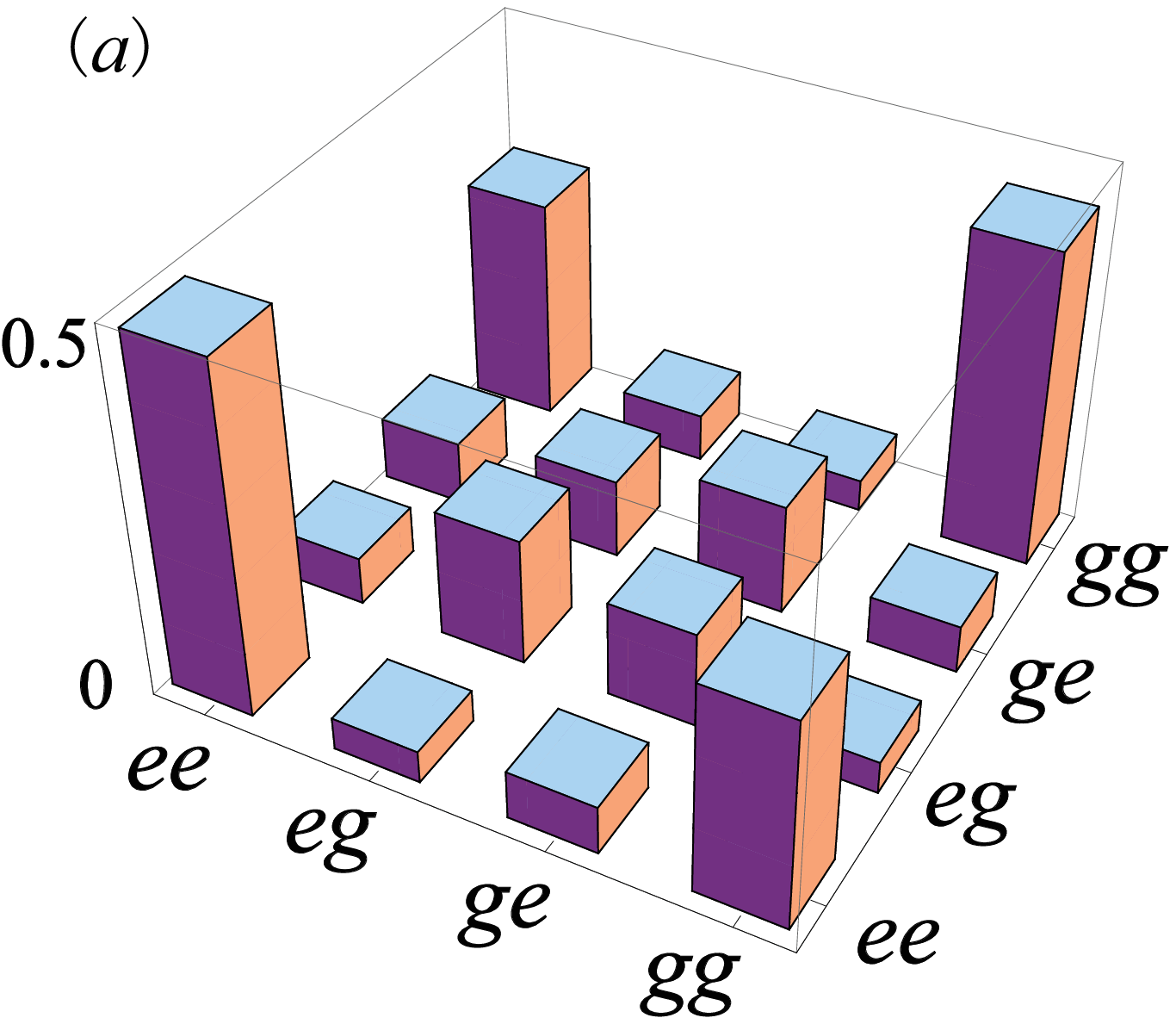}
\includegraphics[width=4.2cm]{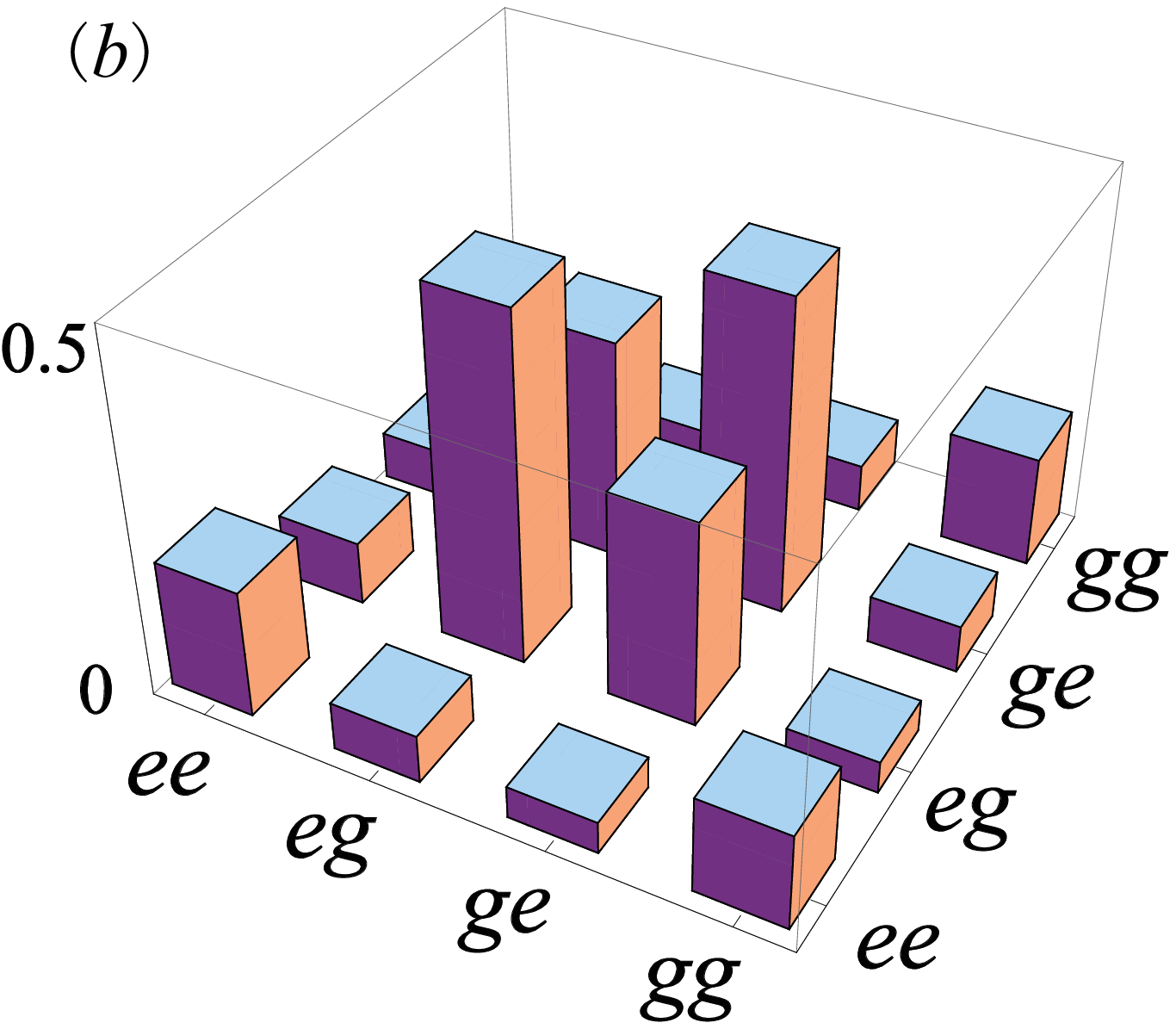}
\caption{Schematic illustration of two types of two-qubit matrices. For
convenience the height of the cuboid represents the modulus of each matrix
element. Plots (a) and (b) illustrate the $\Phi $ type ($Q_{\Phi }\geq
Q_{\Psi }$) and $\Psi $ type ($Q_{\Psi }>Q_{\Phi }$) of two-qubit matrices
respectively.}
\label{matrix}
\end{figure}

\section{Disentanglement Phases for X-form Mixed States}

We now analyze the entanglement dynamics of the X-form initial state $\rho_{X}(0)$, under all three channels, by focusing on the question:
how do initial conditions affect the properties of both CD and TD.

The degree of entanglement, i.e., concurrence \cite{Wootters}, for $\rho
_{X}(0)$ takes a very compact form \cite{Yu-Eberly07},
\begin{equation}
C_{0}=\max \{Q_{\Phi },Q_{\Psi },0\},
\end{equation}
where
\begin{eqnarray}
Q_{\Phi } &=&2(|\rho _{14}|-\sqrt{\rho _{22}\rho _{33}}),  \notag \\
Q_{\Psi } &=&2(|\rho _{23}|-\sqrt{\rho _{11}\rho _{44}}).  \label{Qs}
\end{eqnarray}
There is zero initial entanglement, i.e., $C_{0}=0$, when $Q_{\Phi }<0$ and $Q_{\Psi }<0$. Obviously, this is not an interesting case, and so we will
consider cases when either $Q_{\Phi }$ or $Q_{\Psi }$ is positive (matrix
positivity does not allow both of them to be positive simultaneously). Then
the initial concurrence is expressed conveniently as $C_{0}=Q_{\Phi }$ for $Q_{\Phi }\geq Q_{\Psi }$, or $C_{0}=Q_{\Psi }$ when $Q_{\Phi }<Q_{\Psi}$.

Accordingly, we call the X-form matrix (\ref{X-form initial}) as well as its
generic mother matrix (\ref{general initial}) as a $\Phi $ matrix if $Q_{\Phi }\geq Q_{\Psi }$, or a $\Psi $ matrix otherwise. Fig. \ref{matrix}
gives a schematic illustration of both types of matrices. In practice, a slightly
imperfect preparation of a $\Phi $-like or $\Psi $-like Bell state may lead
to the generation of such a $\Phi $ or $\Psi $ matrix.

\subsection{Amplitude Damping}

For amplitude damping, the time evolution of the initial state $\rho_{X}(0)$ is given in Eq. (\ref{X-amplitude time}). Since it retains the X
form, the time dependent concurrence can be simply obtained as $C(t)=\max
\{Q_{\Phi }(t),Q_{\Psi }(t),0\}$, where
\begin{eqnarray}
Q_{\Phi }(t) &=&2\sqrt{q_{a}q_{b}}(|\rho _{14}|-\sqrt{\Delta (t)+\rho
_{22}\rho _{33}}),  \label{CPhi-amplitude} \\
Q_{\Psi }(t) &=&2\sqrt{q_{a}q_{b}}(|\rho _{23}|-\sqrt{\Delta (t)+\rho
_{11}\rho _{44}}),  \label{CPsi-amplitude}
\end{eqnarray}
and
\begin{equation}
\Delta (t)=\rho _{11}(\rho _{11}p_{a}p_{b}+\rho _{22}p_{a}+\rho _{33}p_{b}).
\end{equation}
The irreversible damping process indicates that the product $q_{a}q_{b}$ is
decreasing while the parameter $\Delta (t)$ is increasing with time. This
guarantees the decay of both $Q_{\Phi }(t)$ and $Q_{\Psi }(t)$. Thus one has
the time dependent concurrence, $C(t)=\max \{Q_{\Phi }(t),0\}$ and $C(t)=\max \{Q_{\Psi }(t),0\}$ for $\Phi $ and $\Psi $ type of initial states
respectively.

In the following two subsections, we analyze the disentanglement properties
of the two ($\Phi $ and $\Psi $) types of X-form initial states
respectively. We set $x_{ij}=\sqrt{\rho _{ii}\rho _{jj}}$ and $y_{ij}=\rho
_{ii}+\rho _{jj}$ with $i,j=1,2,3,4$ for the convenience of notations hereafter.

\subsubsection{$\Phi$ type of X-form initial state}

\begin{figure}[t]
\includegraphics[width=6cm]{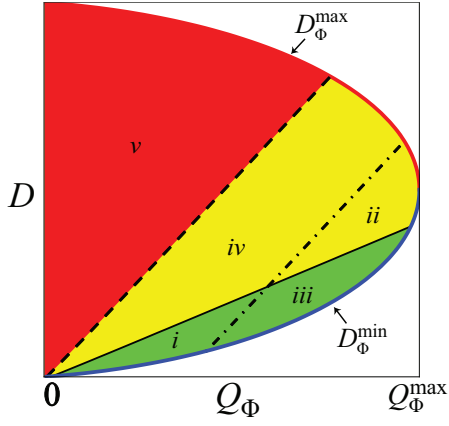}
\caption{Disentanglement phase diagram with respect to double excitation probability
$D$ and entanglement $Q_{\Phi }$ ($C_{0}$) for $\Phi $ type of X-form
initial state under amplitude damping channel. The CD-free critical value $D_{\Phi }^{f}$ is illustrated by the black
solid line. The CD-tolerable ($D_{\Phi }^{t}$) and
TD-tolerable ($D_{\Phi }^{tv}$) critical values are illustrated by the black dashed and
dot-dashed lines respectively for $\gamma _{a}=\gamma _{b}=\gamma $, $\tau =2/3\gamma$, and $C_{tv}=0.1$. The
green zone (regions \textit{i} and \textit{iii}), yellow zone (regions \textit{ii} and \textit{iv}), and red zone (region
\textit{v}) are CD-free (safety), CD-tolerable (caution), and CD-no-go (danger) phases respectively. Regions
\textit{ii} and \textit{iii} form the TD-tolerable phase, and regions \textit{i}, \textit{iv}, \textit{v} form the TD-no-go phase. The overlap region \textit{iii} is the optimal robust phase.}
\label{Phi-PhaseDiagram-amplitude}
\end{figure}

For $\Phi $ type of X-form initial state, of which $C_{0}=Q_{\Phi }$, the
matrix positivity requires $Q_{\Phi }/2\leq x_{14}-x_{23}$. This physical
restriction can be transformed in terms of the double excitation probability
$D\equiv \rho _{11} $, and leads to the relation: $D_{\Phi }^{\min }\leq
D\leq D_{\Phi }^{\max }$, where the physical boundaries are given as
\begin{eqnarray}
2D_{\Phi }^{\min } &=&1-y_{23}-\sqrt{\left( 1-y_{23}\right) ^{2}-\left(
Q_{\Phi }+2x_{23}\right) ^{2}},  \notag \\
2D_{\Phi }^{\max } &=&1-y_{23}+\sqrt{\left( 1-y_{23}\right) ^{2}-\left(
Q_{\Phi }+2x_{23}\right) ^{2}}.  \label{Phi
physical-amplitude}
\end{eqnarray}
Fig. \ref{Phi-PhaseDiagram-amplitude} illustrates the behaviors of $D_{\Phi
}^{\max }$ and $D_{\Phi }^{\min }$ with respect to $Q_{\Phi }$ by the upper
red and lower blue solid lines respectively. The two boundary lines connect
when $D=(1-y_{23})/2$ and $Q_{\Phi }$ takes its maximum value $Q_{\Phi
}^{\max }=1-y_{23}-2x_{23}$.

From (\ref{CPhi-amplitude}), the condition to have CD is obtained as $\Delta (t)+x_{23}^{2}\geq |\rho _{14}|^{2}$. Thus there is no CD
at any time if $\Delta _{\max }(t)+x_{23}^{2}\leq |\rho _{14}|^{2}$, where $\Delta _{\max }(t)$ is the maximum value of $\Delta (t)$ and is achieved
when $p_{a}$, $p_{b}$ take their maximum value 1 (usually correspond to $t=\infty $). This CD-free condition can be re-expressed in terms of the
double excitation as
\begin{equation}
D\leq D_{\Phi }^{f}=\left( \sqrt{y_{23}^{2}+Q_{\Phi }^{2}+4x_{23}Q_{\Phi }}
-y_{23}\right) /2.  \label{ESD
free-Phi--amplitude}
\end{equation}
Here $D_{\Phi }^{f}$ is the CD-free critical value controlled only by the
system initial parameters $Q_{\Phi }$, $\rho _{22}$, and $\rho _{33}$, and
is independent of the time dependent channel characteristic parameters $q_{a} $ and $q_{b}$ that describe reservoir interactions.

The behavior of $D_{\Phi }^{f}$ is illustrated in Fig. \ref{Phi-PhaseDiagram-amplitude} by the black solid line, which crosses the
minimum physical boundary $D_{\Phi }^{\min }$ when $Q_{\Phi }=\sqrt{2x_{23}^{2}+\Lambda }-2x_{23}$, and $2D=\sqrt{y_{23}^{2}-2x_{23}^{2}+\Lambda
}-y_{23}$, with $2\Lambda =1-2y_{23}\pm \sqrt{(1-2y_{23})^{2}-8x_{23}^{2}}$.
From condition (\ref{ESD free-Phi--amplitude}), the region below $D_{\Phi
}^{f}$ in Fig.~\ref{Phi-PhaseDiagram-amplitude} is obviously CD free and everywhere above is CD susceptible. One
can easily imagine that these two CD phases are manageable by only
adjusting the initial parameters $D$, $\rho _{22}$, $\rho _{33}$, and $Q_{\Phi }$.

We further analyze the CD onset condition, i.e., $Q_{\Phi }(t)=0$. A typical conventional approach is to solve this equation for $t$ to determine the CD time $t_{CD}$ by taking the single qubit channel dissipation, for example, as an exponential decay process, i.e., $p_{k}=p_{k}(t)=1-e^{-\gamma _{k}t}$ with $\gamma _{k}$ ($k=a,b$) as the decay constants for qubits $A$, $B$ respectively. Then the CD time can be determined as $t_{CD}=\ln{s}$, where $s$ is the real positive solution of the linear equation
\begin{eqnarray}
0&=&[ \rho_{11}(1- \rho_{44}) +\rho_{22}\rho_{33}- |\rho_{14}|^2 ] s^{\gamma_{a}+\gamma_{b}}  \notag \\
 && -\rho_{11}^2 - \rho_{11}\rho_{22} s^{\gamma_{a}} -
 \rho_{11}\rho_{33} s^{\gamma_{b}}.  \label{ESD time difficult}
\end{eqnarray}
Obviously, such a linear equation can be solved for symmetric dissipation channels, i.e., $\gamma _{a}=\gamma _{b}$. However for any non-symmetric cases, i.e., $\gamma _{a}\neq\gamma _{b}$, it will be extremely complicated if not impossible to achieve analytical solutions.

Here we consider practical situations when a quantum information task can tolerate CD if it
happens after a prescribed time $\tau $ that is determined by intrinsic computation and operation time scales. CD that happens earlier than $\tau $ may be considered as harmful and is a no-go region. Then we are able to avoid the difficulty to obtain CD onset time by taking the inverse approach, i.e., solving the initial conditions for arbitrary excitation-transfer probabilities $p_{k}(\tau)$ at a given time $\tau$. In this way the CD-tolerable condition $t_{CD}\geq \tau $ can be expressed in terms of the double excitation $D$ bounded by its critical
value $D_{\Phi }^{t}$, i.e., $D\leq D_{\Phi }^{t}$ with
\begin{eqnarray}
D_{\Phi }^{t} &=&\sqrt{\left( \frac{\rho _{22}}{2p_{b}(\tau )}-\frac{\rho
_{33}}{2p_{a}(\tau )}\right) ^{2}+\frac{(Q_{\Phi }+2x_{23})^{2}}{4p_{a}(\tau
)p_{b}(\tau )}}  \notag \\
&&-\frac{\rho _{22}}{2p_{b}(\tau )}-\frac{\rho _{33}}{2p_{a}(\tau )}.\label{ESD-tolerable Phi-amplitude}
\end{eqnarray}
One notes that $D_{\Phi
}^{t}$ is also controlled by the initial system parameters $\rho _{22}$, $\rho _{33}$, and $Q_{\Phi }$ when the tolerable time scale $\tau $ is given. Note that $p_{a}(\tau )=0$ and $p_{b}(\tau )=0$ are singularities of $D_{\Phi }^{t}$ and they represent single-channel dissipation cases. The disentanglement properties of those situations are similar to the
two-channel case and will not be addressed in detail here.

For demonstration, we also take the usual exponential decay assumption, i.e., $p_{k}(t)=1-e^{-\gamma _{k}t}$. Then the behavior of $D_{\Phi }^{t}$ is illustrated in Fig. \ref{Phi-PhaseDiagram-amplitude} by the dashed line with $\gamma _{a}=\gamma _{b}=\gamma $ (just for convenience) and $\tau=2/3\gamma$. In Fig. \ref{Phi-PhaseDiagram-amplitude}, the green zone (regions \textit{i }
and \textit{iii}) is the CD-free phase, and is considered as the safety
phase with $t_{CD}=\infty $. One notes that $D_{\Phi }^{t}$ splits the CD
susceptible area into two zones. The yellow zone (regions \textit{ii }and
\textit{iv}) is the CD-tolerable
(caution) phase in which CD occurs only after the tolerable time scale, i.e., $t_{CD}\geq \tau $. The red zone (region \textit{v}) is the CD-no-go (danger) phase in which CD happens before $\tau$, where
state preparations should avoid. We remark that the situations when $\gamma
_{a}$, $\gamma _{b}$ and $\tau $ are arbitrary will give qualitatively
similar behaviors as predicted by Eq.~(\ref{ESD-tolerable Phi-amplitude}),
and the yellow CD-tolerable region will increase with the decrease of $\gamma
_{a}$, $\gamma _{b}$ and $\tau $.

For a complete illustration of disentanglement management, we consider further another important practical situation, i.e., some quantum
information schemes may require the existing entanglement (e.g., in terms of concurrence) to stay above a non-zero threshold value $C_{tv}$ before the tolerable time $\tau$. We call this a threshold disentanglement process, and the condition to meet this requirement is
$Q_{\Phi }(\tau )\geq C_{tv}$, where $Q_{\Phi }(\tau )$ is given in Eq. (\ref{CPhi-amplitude}) for time $\tau$. This relation can be simply transformed into an inequality in which the initial double excitation $D$ is bounded by a
critical value $D_{\Phi }^{tv}$, i.e., $D\leq D_{\Phi }^{tv}$ with
\begin{eqnarray}
D_{\Phi }^{tv} &=&\sqrt{\left( \frac{\rho _{22}}{2p_{b}(\tau )}-\frac{\rho
_{33}}{2p_{a}(\tau )}\right) ^{2}+\frac{(Q_{\Phi }+2x_{23}-\delta )^{2}}{4p_{a}(\tau )p_{b}(\tau )}} \notag \\
&&-\frac{\rho _{22}}{2p_{b}(\tau )}-\frac{\rho _{33}}{2p_{a}(\tau )}.
\end{eqnarray}
Here $\delta =C_{tv}/\sqrt{q_{a}(\tau )q_{b}(\tau )}$ is a shift term
comparing to the CD-tolerable critical value $D_{\Phi }^{t}$. One notes
that $D_{\Phi }^{tv}$ is controlled by the initial system parameters $\rho
_{22}$, $\rho _{33}$, and $Q_{\Phi }$ in the same fashion as $D_{\Phi }^{t}$. The behavior of $D_{\Phi }^{tv}$ is illustrated by the dot-dashed line in Fig. \ref{Phi-PhaseDiagram-amplitude} with $\gamma _{a}=\gamma _{b}=\gamma $, $\tau =2/3\gamma $ and $C_{tv}=0.1$. Regions \textit{ii}, \textit{iii} form the TD-tolerable phase with the time $t_{tv}$ to reach threshold value $C_{tv}$ longer than the tolerable time scale, i.e., $t_{tv}\geq \tau $, and regions \textit{i}, \textit{iv},
\textit{v} form the TD-no-go phase with $t_{tv}<\tau $.

\begin{figure}[t]
\includegraphics[width=5cm]{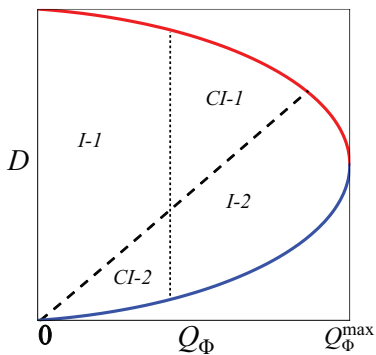}
\caption{The first type of intuitive and counter-intuitive CD regions with respect to initial entanglement $C_{0}$ ($Q_{\Phi }$). The black dashed line represents the CD-tolerable critical value $D_{\Phi }^{t}$ and crosses with a vertical dotted line. The states of regions I-1 and CI-1 are in the CD-no-go phase with relatively earlier CD than the states in CD-tolerable phase, i.e., regions I-2 and CI-2. I-1 and I-2 form a pair of intuitive regions, and CI-1 and CI-2 form a pair of counter-intuitive regions.}
\label{counter-intuitive}
\end{figure}

There are two types of intuitive and counter-intuitive disentanglement situations. The first type concerns CD (and TD) properties with respect to initial entanglement $C_{0}$ ($Q_{\Phi }$). As shown in Fig.~\ref{counter-intuitive}, regions I-1 and I-2 are a pair of intuitive CD regions where states with less (more) initial entanglement decays to zero earlier (later), and regions CI-1 and CI-2 are a pair of counter-intuitive CD regions where states with less (more) $Q_{\Phi }$ reach CD later (earlier). Similarly, the intuitive and counter-intuitive regions of TD can also be identified with a vertical line that crosses with the TD-tolerable critical line $D_{\Phi }^{tv}$. The second type concerns CD properties with respect to TD properties, and is shown explicitly in Fig.~\ref{Phi-PhaseDiagram-amplitude}. The counter-intuitive states suffer earlier CD (region \textit{ii}) reach the threshold value later than those with longer CD time (region \textit{i}), and the intuitive situation where states are relatively more robust against both CD and TD is shown by region \textit{iii}. Thus region \textit{iii} is recognized as the optimal robust phase.

From the above analysis, we conclude for $\Phi $ type X-form initial state under amplitude damping noise, that CD-free, CD-tolerable, CD-no-go, TD-tolerable, TD-no-go, and optimal robust phases are fully manageable through the initial system parameters $D$, $\rho _{22}$, $\rho _{33}$, and $Q_{\Phi }$ when the tolerable time scale $\tau $ is given.

\subsubsection{$\Psi$ type of X-form initial state}

\begin{figure}[t]
\includegraphics[width=6cm]{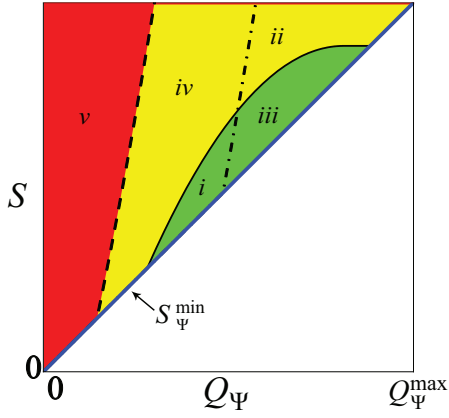}
\caption{Disentanglement phase diagram with respect to single
excitation probability $S$ and entanglement $Q_{\Psi }$ ($C_{0}$) for $\Psi $
type of X-form initial state under amplitude
damping channel. The CD-free critical value $S_{\Psi }^{f}$ is illustrated by the
black solid line. The CD-tolerable ($S_{\Psi }^{t}$)
and TD-tolerable ($S_{\Psi }^{tv}$) critical values are illustrated by the black dashed and
dot-dashed lines respectively for $\gamma _{a}=\gamma _{b}=\gamma $, $\tau =2/3\gamma $ and $C_{tv}=0.1$. The
green zone (regions \textit{i} and \textit{iii}), yellow zone
(regions \textit{ii} and \textit{iv}), and red zone (region
\textit{v}), are CD-free (safety), CD-tolerable (caution), and CD-no-go (danger) phases respectively. Regions
\textit{ii} and \textit{iii} form the TD-tolerable
phase, and regions \textit{i}, \textit{iv }, \textit{v} form the TD-no-go phase. The overlap region \textit{iii} is the
optimal robust phase.}
\label{Psi-PhaseDiagram-amplitude}
\end{figure}

For $\Psi $ type of X-form matrix, one has the initial concurrence $C_{0}=Q_{\Psi }$. The matrix positivity requires the relation $Q_{\Psi
}/2\leq x_{23}-x_{14}$. This can be transformed in terms of the single
excitation probability $S=\rho _{22}=\rho _{33}$, and leads to the
restriction: $S_{\Psi}^{\min }\leq S\leq 1/2$, where the minimum physical
boundary is given as
\begin{equation}
S_{\Psi}^{\min }=Q_{\Psi }/2+\sqrt{\rho _{11}\rho _{44}}.
\label{Psi
physical-amplitude}
\end{equation}
Here, without loss of generality, we have assumed $\rho _{22}=\rho _{33}$
for the convenience of our analysis. Fig.~\ref{Psi-PhaseDiagram-amplitude}
illustrates the behavior of $S_{\Psi}^{\min }$ by the blue solid line as a
function of $Q_{\Psi }$. Along this boundary, $Q_{\Psi }$ takes its maximum
value $Q_{\Psi }^{\max }=1-2x_{14}$ when $S=1/2$.

From Eq.~(\ref{CPsi-amplitude}), one notes that the condition to have CD is
whenever $\Delta (t)+x_{14}^{2}\geq |\rho _{23}|^{2}$. Thus the CD-free
condition is achieved as $\Delta _{\max }(t)+x_{14}^{2}\leq |\rho _{23}|^{2}$, which can be re-expressed in terms of the single excitation probability bounded by its critical value $S_{\Psi }^{f}$, i.e.,
\begin{equation}
S\leq S_{\Psi }^{f}=\frac{4Q_{\Psi }\sqrt{\rho _{11}}-Q_{\Psi }^{2}-4\rho
_{11}^{2}}{8\rho _{11}}.  \label{ESD free-Psi-amplitude}
\end{equation}
Here $S_{\Psi }^{f}$ depends only on the system initial parameters $Q_{\Psi
} $, $\rho _{11}$, see illustration in Fig.~\ref{Psi-PhaseDiagram-amplitude} by the black solid line. Its crossing with the
physical boundary $S_{\Psi }^{\min }$ can be obtained straightforwardly through
the equation $S_{\Psi }^{f}=S_{\Psi }^{\min }$. The green zone (regions \textit{i }and
\textit{iii}) is the CD-free (safety) phase with $t_{CD}=\infty $ and everywhere else is CD susceptible.

The CD onset relation is given as $Q_{\Psi }(t)=0$. Again, it will be very difficult to determine analytically the CD time $t_{CD}$ in most cases as shown by Eq.~(\ref{ESD time difficult}). Thus we solve the inverse problem and express the CD-tolerable condition $t_{CD}\geq \tau $ in terms of the initial single excitation probability as $S\leq S_{\Psi }^{t}$, where the CD-tolerable critical value
$S_{\Psi }^{t}$ is given as
\begin{equation}
S_{\Psi }^{t}=\frac{(Q_{\Psi }+2x_{14})^{2}-4x_{14}^{2}-4\rho
_{11}^{2}p_{a}(\tau )p_{b}(\tau )}{4\rho _{11}\left[ p_{a}(\tau )+p_{b}(\tau
)\right] }  \label{ESD-tolerable Psi-amplitude}
\end{equation}
Obviously, it is controlled by the initial parameters $Q_{\Psi }$, $\rho
_{11}$, and $\rho _{44}$ when the tolerable time scale $\tau $ is fixed. To
illustrate $S_{\Psi }^{t}$, we make the same assumption $p_{k}(t)=1-e^{-\gamma _{k}t}$ with $k=a,$ $b$, and take $\gamma _{a}=\gamma _{b}=\gamma $, $\tau =2/3\gamma $ for convenience. The black dashed line in Fig. \ref{Psi-PhaseDiagram-amplitude} represents the behavior of $S_{\Psi }^{t}$. It
splits the CD-susceptible zone into two, of which the yellow zone
(regions \textit{ii }and \textit{iv}) is the CD-tolerable (caution) phase
with $t_{CD}\geq \tau $ and the red zone is the CD-no-go (danger) phase
with $t_{CD}<\tau $. Again, situations for arbitrary $\gamma _{a}$, $\gamma
_{b}$, and $\tau $ are determined by (\ref{ESD-tolerable Psi-amplitude}), and
these will have similar phase diagrams, of which the yellow CD-tolerable
region will increase with the decrease of $\gamma _{a}$, $\gamma _{b}$, and $\tau$.

We further consider threshold disentanglement, where the TD-tolerable
states are decided by the relation $Q_{\Psi }(\tau )\geq C_{tv}$. Again this relation is transformed into the restriction $S\leq S_{\Psi }^{tv}$, where
\begin{equation}
S_{\Psi }^{tv}=\frac{(Q_{\Psi }+2x_{14}-\delta )^{2}-4x_{14}^{2}-4\rho
_{11}^{2}p_{a}(\tau )p_{b}(\tau )}{4\rho _{11}\left[ p_{a}(\tau )+p_{b}(\tau
)\right] }
\end{equation}
is the TD critical value for single excitation probability with a shift of $\delta =C_{tv}/\sqrt{q_{a}(\tau )q_{b}(\tau )}$ comparing to the CD
tolerable critical value $S_{\Psi }^{t}$. The behavior of $S_{\Psi }^{tv}$
is illustrated by the dot-dashed line in Fig. \ref{Psi-PhaseDiagram-amplitude} with $\gamma _{a}=\gamma _{b}=\gamma $, $\tau
=2/3\gamma $ and $C_{tv}=0.1$. Regions \textit{ii}, \textit{iii} form the TD-tolerable phase with $t_{tv}\geq \tau $, and regions \textit{i}, \textit{iv}, \textit{v} form the TD-no-go phase with $t_{tv}<\tau $.

Again, the first type of intuitive and counter-intuitive CD and TD situations can be identified easily by comparing different regions separated by the crossing of a vertical line with various disentanglement critical lines (similar to Fig.~\ref{counter-intuitive}). The second type is explicitly shown in Fig.~\ref{Psi-PhaseDiagram-amplitude}, where \textit{i} and \textit{ii} are a pair of counter-intuitive regions where states suffer earlier (later) TD reaches CD later (earlier). Region \textit{iii} is the intuitive optimal robust phase.

Consequently, CD-free, CD-tolerable, CD-no-go, TD-tolerable,
TD-no-go, and optimal robust phases of the $\Psi$ type X-form initial states are controllable
through the initial system parameters $S$, $\rho _{11}$, $\rho _{44}$, and $Q_{\Psi }$.

As an intermediate conclusion, we have shown for amplitude damping
noise, that various disentanglement phases can be managed through just a few
system initial parameters for X-form states.

\subsection{Phase Damping}

The time dependent state of the X-form initial state $\rho_{X}(0)$
for phase damping channel is given in Eq.~(\ref{X-phase time}). The time dependent concurrence can be simply obtained
as $C(t)=\max \{Q_{\Phi }(t),Q_{\Psi }(t),0\}$, with
\begin{eqnarray}
Q_{\Phi }(t) &=&2(|\rho _{14}\sqrt{q_{a}q_{b}}|-\sqrt{\rho _{22}\rho _{33}})
\label{CPhi-phase} \\
Q_{\Psi }(t) &=&2(|\rho _{23}\sqrt{q_{a}q_{b}}|-\sqrt{\rho _{11}\rho _{44}})
\label{CPsi-phase}
\end{eqnarray}
The irreversible decay of the product $q_{a}q_{b}$ guarantees the decreasing
of both $Q_{\Phi }(t)$ and $Q_{\Psi }(t)$. Thus one has the time dependent
concurrence, $C(t)=\max \{Q_{\Phi }(t),0\}$ and $C(t)=\max \{Q_{\Psi
}(t),0\} $ for $\Phi $ and $\Psi $ type of initial matrices respectively. We
analyze both situations explicitly in the following two subsections.

\subsubsection{$\Phi $ type of X-form initial state}

\begin{figure}[t]
\includegraphics[width=6cm]{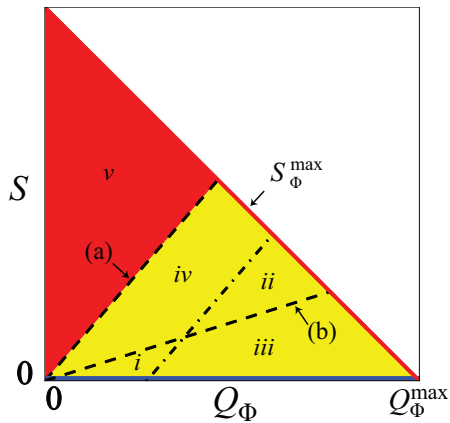}
\caption{Disentanglement phase diagram with respect to single
excitation probability $S$ and entanglement $Q_{\Phi }$ ($C_{0}$) for $\Phi$
type of X-form initial state under phase damping channel. The blue solid line $S_{\Phi }^{f}=0$ denotes the CD-free
phase. The CD-tolerable critical value $S_{\Phi }^{t}$ is illustrated by
the dashed lines (a) and (b) for $\tau $ takes $1/\gamma $
and $2/\gamma $ respectively. The TD-tolerable critical value $S_{\Phi }^{tv}$ is illustrated by the dot-dashed line for $\tau =1/\gamma$ and $C_{tv}=0.1$. For $\tau =1/\gamma$, the yellow zone (regions \textit{i},
\textit{ii}, \textit{iii} and \textit{iv}) and the red zone (region
\textit{v}) are CD-tolerable (caution) and CD-no-go (danger) phases respectively. Regions
\textit{ii} and \textit{iii} form the TD-tolerable
phase, regions \textit{i}, \textit{iv }, \textit{v} form the TD-no-go phase, and region \textit{iii} is the optimal robust phase.}
\label{Phi-PhaseDiagram-phase}
\end{figure}

For $\Phi $ type of X-form matrix, the initial concurrence is simply given
as $C_{0}=Q_{\Phi }$, and the matrix positivity requires $Q_{\Phi }/2\leq
x_{14}-x_{23}$. We rewrite this physical restriction in terms of the single
excitation probability $S=\rho _{22}=\rho _{33}$, as $0\leq S\leq S_{\Phi
}^{\max }$, where the maximum boundary is given as
\begin{equation}
S_{\Phi }^{\max }=(1-Q_{\Phi })/4.  \label{Phi physical-phase}
\end{equation}
Again, we have assumed $\rho _{22}=\rho _{33}$ without loss of generality.
The behavior of $S_{\Phi }^{\max }$ is illustrated by the red solid line in
Fig.~\ref{Phi-PhaseDiagram-phase}. Along this physical boundary line, $Q_{\Phi }$ takes its minimum value $Q_{\Phi }^{\min }=0$ when $S_{\Phi
}^{\max }=1/4$, and reaches its maximum value $Q_{\Phi }^{\max }=1$ when $S_{\Phi }^{\max }=0$.

From (\ref{CPhi-phase}), one immediately finds that CD free happens only
when $\rho _{22}\rho _{33}=0$, i.e., $S=0$. Therefore the
CD-free region with $t_{CD}=\infty $ shrinks to just a line as illustrated
by the blue solid line $S_{\Phi }^{f}=0$ in Fig.~\ref{Phi-PhaseDiagram-phase}, and everywhere else is CD susceptible.

The condition to have CD is whenever $|\rho _{14}\sqrt{q_{a}q_{b}}|=\sqrt{\rho _{22}\rho _{33}}$. One notes from (\ref{CPhi-phase}) that for phase damping channel CD onset time can be determined analytically. For the purpose of the current paper we will consider the inverse initial condition problem. The CD tolerable condition $t_{CD}\geq \tau$, i.e., $|\rho _{14}\sqrt{q_{a}(\tau )q_{b}(\tau )}|\geq \sqrt{\rho _{22}\rho _{33}}$, can be transformed in terms of the single excitation probability bounded by its
critical value $S_{\Phi }^{t}$, i.e.,
\begin{equation}
S\leq S_{\Phi }^{t}=\frac{Q_{\Phi }\sqrt{q_{a}(\tau )q_{b}(\tau )}}{2[1-\sqrt{q_{a}(\tau )q_{b}(\tau )}]}.  \label{ESD-tolerable
Phi-phase}
\end{equation}
Obviously, $S_{\Phi }^{t}$ is controlled by the initial entanglement $Q_{\Phi }$ for any fixed tolerable time scale $\tau$.

To illustrate various disentanglement phases, we again assume $p_{k}(t)=1-e^{-\gamma _{k}t}$ with $k=a,b$ and take $\gamma _{a}=\gamma
_{b}=\gamma $ for convenience. The behavior of $S_{\Phi }^{t}$ is
illustrated in Fig. \ref{Phi-PhaseDiagram-phase} by the black dashed lines (a) and (b) for $\tau $ equals $1/\gamma $ and $2/\gamma$ respectively. Taking $\tau=1/\gamma $ as an example, the yellow zone (regions \textit{i}, \textit{ii},
\textit{iii}, and \textit{iv}) is the CD-tolerable phase with $t_{CD}\geq \tau $, and the
red zone (region \textit{v}) is the CD-no-go phase with $t_{CD}<\tau $.
Again, the yellow CD-tolerable region will decrease with the increase of $\gamma _{a}$, $\gamma
_{b}$ and $\tau$, see for example a lager value of $\tau=2/\gamma$ indicated by line (b). The crossing of
the CD-tolerable critical line with the maximum physical boundary line can
be simply computed by solving the equation $S_{\Phi }^{t}=S_{\Phi }^{\max }$.

We now consider the threshold disentanglement situation. The TD-tolerable
requirement is determined by the relation $Q_{\Phi }(\tau )\geq C_{tv}$. Again this relation can be transformed into the restriction
\begin{equation}
S\leq S_{\Phi }^{tv}=\frac{(Q_{\Phi }-\delta )\sqrt{q_{a}(\tau )q_{b}(\tau )}}{2[1-\sqrt{q_{a}(\tau )q_{b}(\tau )}]},
\end{equation}
where $S_{\Phi }^{tv}$ is the threshold critical value with a shift of $\delta =C_{tv}/\sqrt{q_{a}(\tau )q_{b}(\tau )}$
comparing to the CD tolerable critical value $S_{\Phi }^{t}$, and it is
also controlled by the initial entanglement $Q_{\Phi }$. The behavior
of $S_{\Phi }^{tv}$ illustrated by the dot-dashed line in Fig.~\ref{Phi-PhaseDiagram-phase} for $\gamma _{a}=\gamma _{b}=\gamma $, $\tau
=1/\gamma $ and $C_{tv}=0.1$. Regions \textit{ii}, \textit{iii} form the TD-tolerable phase with $t_{tv}\geq \tau $, and
regions \textit{i}, \textit{iv }, \textit{v} form the TD-no-go phase with $t_{tv}<\tau$.

The first type of intuitive and counter-intuitive CD and TD situations can be identified by comparing different regions separated by the crossing of a vertical line with various disentanglement critical lines (similar to Fig.~\ref{counter-intuitive}). The second type is shown in Fig.~\ref{Phi-PhaseDiagram-phase}. Regions \textit{i} and \textit{ii} that are separated by the crossing of $S_{\Phi }^{tv}$ with line (b) illustrate the counter-intuitive situation, i.e., sates in region \textit{ii} suffer earlier CD but reach TD later than the sates in region \textit{i}. Region \textit{iii} is the intuitive optimal robust phase against both CD and TD.

Thus we have shown for $\Phi $ type of X-form initial state under phase
damping channel, that various CD and TD phases are manageable through the
initial parameters $S$, and $Q_{\Phi }$ for any fixed tolerable time scale $\tau $.

\subsubsection{$\Psi$ type of X-form initial state}

\begin{figure}[t]
\includegraphics[width=6cm]{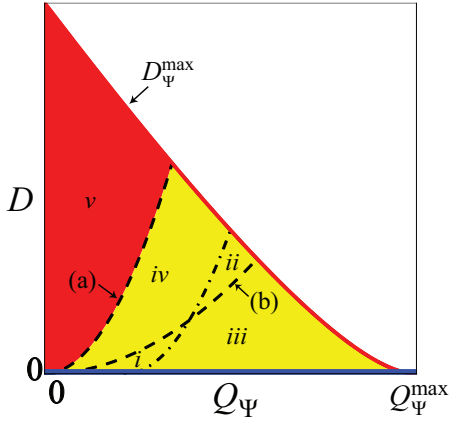}
\caption{Disentanglement phase diagram with respect to double
excitation probability $D$ and entanglement $Q_{\Psi }$ ($C_{0}$) for $\Psi$
type of X-form initial state under phase damping channel. The blue solid line $D_{\Psi }^{f}=0$ denotes the CD-free
phase. The CD-tolerable critical value $D_{\Psi }^{t}$ is illustrated by
the dashed lines (a) and (b) for $\tau $ takes $2/3\gamma $
and $8/7\gamma $ respectively. The TD-tolerable critical value $D_{\Psi }^{tv}$ is illustrated by the dot-dashed line for $\tau =2/3\gamma $ and $C_{tv}=0.1$. For $\tau =2/3\gamma$, the yellow zone (regions \textit{i},
\textit{ii}, \textit{iii} and \textit{iv}) and red zone (region
\textit{v}) are CD-tolerable (caution) and CD-no-go (danger) phases respectively. Regions
\textit{ii} and \textit{iii} form the TD-tolerable
phase, regions \textit{i}, \textit{iv }, \textit{v} form the TD-no-go phase, and region \textit{iii} is the optimal robust phase.}
\label{Psi-PhaseDiagram-phase}
\end{figure}

For the $\Psi $ matrix situation, i.e., $C_{0}=Q_{\Psi }$, the matrix
positivity requires $Q_{\Psi }/2\leq x_{23}-x_{14}$. This can be transformed
in terms of the double excitation probability $D\equiv \rho _{11}$, i.e., $0\leq D\leq D_{\Psi }^{\max }$, where the maximum physical boundary is given as
\begin{equation}
D_{\Psi }^{\max }=(\sqrt{1-Q_{\Psi }}-\sqrt{\rho _{44}})^{2}\text{.}
\label{Psi physical-phase}
\end{equation}
Fig.~\ref{Psi-PhaseDiagram-phase} illustrates the behavior of $D_{\Psi
}^{\max }$ explicitly as a function of $Q_{\Psi }$ by the red solid line.
Along this boundary line $Q_{\Psi }$ takes the maximum value $Q_{\Psi
}^{\max }=1-\rho _{44}$ when $D_{\Psi }^{\max }=0$, and reaches the minimum
value $Q_{\Psi }^{\min }=0$ when $D_{\Psi }^{\max }=(1-\sqrt{\rho _{44}})^{2} $.

Eq.~(\ref{CPsi-phase}) indicates that there is no CD when $\rho _{11}\rho _{44}=0$,
i.e., $\rho _{11}=0$ or $\rho _{44}=0$. Here we consider the CD-free phase in
terms of the double excitation probability, and it is just a single line $D_{\Psi
}^{f}=0$ as illustrated by the blue solid line in Fig.~\ref{Psi-PhaseDiagram-phase}.

The CD tolerable condition $t_{CD}\geq \tau$ is
determined by the relation $|\rho _{23}\sqrt{q_{a}(\tau )q_{b}(\tau )}|\geq
\sqrt{\rho _{11}\rho _{44}}$, which can also be transformed in terms of the
double excitation probability bounded by its CD-tolerable critical value $D_{\Psi }^{t}$, i.e.,
\begin{equation}
D\leq D_{\Psi }^{t}=\frac{Q_{\Psi }^{2}q_{a}(\tau )q_{b}(\tau )}{4[1-\sqrt{q_{a}(\tau )q_{b}(\tau )}]^{2}\rho _{44}}.  \label{ESD-tolerable Psi-phase}
\end{equation}
It is controlled by the initial system parameters $Q_{\Psi }$ and $\rho
_{44} $ for a given tolerable time scale $\tau $.

For demonstration, we make the same exponential decay assumption, i.e., $p_{k}(t)=1-e^{-\gamma _{k}t}$ with $k=a,b$, and take $\gamma _{a}=\gamma
_{b}=\gamma $ for convenience. Then the exact behavior of $D_{\Psi }^{t}$ is
illustrated in Fig. \ref{Psi-PhaseDiagram-phase} by the dot-dashed lines (a)
and (b) for $\tau =2/3\gamma $ and $\tau =8/7\gamma $ respectively. Taking $\tau =2/3\gamma$ as an example,
the yellow zone (regions \textit{i}, \textit{ii}, \textit{iii}, and \textit{iv}) is the CD-tolerable phase with $t_{CD}\geq \tau $, and the red zone (region \textit{v}) is the CD-no-go phase with $t_{CD}<\tau $. The yellow CD-tolerable region
will decrease with the increase of $\gamma _{a}$, $\gamma _{b}$ and $\tau $, see for example a lager
value of $\tau =8/7\gamma $ indicated by line (b). The crossing of the CD-tolerable critical line and
the maximum physical boundary line can be simply computed by solving the
equation $D_{\Psi }^{t}=D_{\Psi }^{\max }$.

We further consider the threshold disentanglement case, where the
TD-tolerable requirement is given as $Q_{\Psi }(\tau )\geq C_{tv}$, and it
can be transformed into a restriction of the double excitation probability
\begin{equation}
D\leq D_{\Psi }^{tv}=\frac{(Q_{\Psi }-\delta )^{2}q_{a}(\tau )q_{b}(\tau )}{4[1-\sqrt{q_{a}(\tau )q_{b}(\tau )}]^{2}\rho _{44}}.
\label{TD-tolerable Psi-phase}
\end{equation}
Here $D_{\Psi }^{tv}$ is the threshold critical value with a shift of $\delta =C_{tv}/\sqrt{q_{a}(\tau )q_{b}(\tau )}$ comparing to the CD
tolerable critical value $D_{\Psi }^{t}$, and it is also only controlled by
the initial entanglement $Q_{\Psi }$ and $\rho _{44}$. The behavior of $D_{\Psi }^{tv}$ is illustrated by the dot-dashed line in Fig. \ref{Psi-PhaseDiagram-phase} for $\gamma _{a}=\gamma _{b}=\gamma $, $\tau
=2/3\gamma $ and $C_{tv}=0.1$. Regions \textit{ii}, \textit{iii}
form the TD-tolerable phase with $t_{tv}\geq \tau $ and regions \textit{i}, \textit{iv}, \textit{v} form the
TD-no-go phase with $t_{tv}<\tau $.

Again, the first type of intuitive and counter-intuitive disentanglement situations can be identified easily by comparing different regions separated by the crossing of a vertical line with various disentanglement critical lines (similar to Fig.~\ref{counter-intuitive}). The second type is shown in Fig.~\ref{Psi-PhaseDiagram-phase}, where the counter-intuitive situation is identified by comparing regions \textit{i} and \textit{ii} that are separated by the crossing of $D_{\Psi }^{tv}$ with line (b). Obviously, states in region \textit{ii} suffer earlier CD but experience later TD than states in region \textit{i}. Again region \textit{iii} is the intuitive optimal robust phase against both CD and TD.

From the above analysis, we conclude for $\Psi $ type of X-form initial
state under phase damping channel, that various CD and TD phases are manageable through the
initial parameters $D$, $Q_{\Psi }$ and $\rho _{44}$ for any given tolerable
time scale $\tau $.

To this end, we have shown that the pre-management of disentanglement properties of an X-form state is also feasible under phase damping channel.

\subsection{Depolarization}

The time dependent state of the initial X-form state $\rho_{X}(0)$
under depolarization channel is given in Eq.~(\ref{X-polarization time}). The time dependent concurrence can be
simply obtained as $C(t)=\max \{Q_{\Phi }(t),Q_{\Psi }(t),0\}$, with
\begin{eqnarray}
Q_{\Phi }(t) &=&2[|\rho _{14}(t)|-\sqrt{\rho _{22}(t)\rho _{33}(t)}],
\label{CPhi-depolarization} \\
Q_{\Psi }(t) &=&2[|\rho _{23}(t)|-\sqrt{\rho _{11}(t)\rho _{44}(t)}],
\label{CPsi-depolarization}
\end{eqnarray}
One then has $C(t)=\max \{Q_{\Phi }(t),0\}$ and $C(t)=\max \{Q_{\Psi }(t),0\}$ for $\Phi $ and $\Psi
$ type of initial matrices respectively.

CD-free condition requires that there is a region of initial conditions within which $Q_{\Phi }(t)>0$ for
any $t<\infty $. We check the limiting case when $t\rightarrow \infty $, and
find that $Q_{\Phi }(t)\rightarrow -1/2$, which obviously indicates that $C_{\Phi }(t)=0$ at finite time. That is, there is no CD-free region for $\Phi $ type of initial matrices under depolarization channel. Similarly, it can be shown that $\Psi $ type of matrix doesn't contain an CD free-region either. Thus, in the following two subsections we analyze the CD-tolerable,
CD-no-go, TD-tolerable, and TD-no-go phases for $\Phi $ and $\Psi $ type of initial states respectively.

\subsubsection{$\Phi $ type of X-form initial state}

\begin{figure}[t]
\includegraphics[width=6cm]{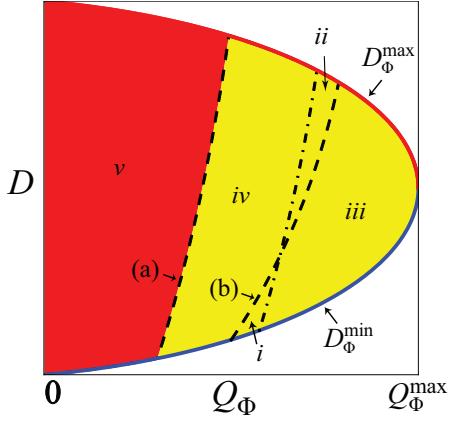}
\caption{Disentanglement phase diagram with respect to double
excitation probability $D$ and entanglement $Q_{\Phi }$ ($C_{0}$) for $\Phi$
type of X-form initial state under
depolarization channel. The CD-tolerable critical value $D_{\Phi }^{t}$ is
illustrated by the dashed lines (a) and (b) for $\tau $ takes $1/5\gamma $ and $11/36\gamma $ respectively. The
TD-tolerable critical value $D_{\Phi }^{tv}$ is illustrated by the
dot-dashed line for $\tau =1/5\gamma $ and $C_{tv}=0.1$. For $\tau =1/5\gamma$, the
yellow zone (regions \textit{i}, \textit{ii}, \textit{iii} and \textit{iv}) and red zone (region \textit{v}) are CD-tolerable (caution) and CD-no-go (danger) phases respectively. Regions \textit{ii} and \textit{iii} form the TD-tolerable phase, regions \textit{i}, \textit{iv}, \textit{v}
form the TD-no-go phase, and region \textit{iii} is the
optimal robust phase.}
\label{Phi-PhaseDiagram-polarization}
\end{figure}

For $\Phi $ type of X-form initial state, one has $C_{0}=Q_{\Phi }$. The
matrix positivity requires $Q_{\Phi }/2\leq x_{14}-x_{23}$. Again we can
transform this restriction in terms of the double excitation probability, i.e., $D_{\Phi }^{\min }\leq D\leq D_{\Phi }^{\max }$,
where the physical boundaries $D_{\Phi }^{\min }$ and $D_{\Phi }^{\max }$ are the same as given in (\ref{Phi physical-amplitude}).
The behaviors of these two boundaries are illustrated again in Fig.~\ref{Phi-PhaseDiagram-polarization} by
the upper red and lower blue solid lines respectively.

From (\ref{CPhi-depolarization}), the CD onset condition can be obtained as
$|\rho _{14}(t)|=\sqrt{\rho _{22}(t)\rho _{33}(t)}$. Similar to the reason shown in (\ref{ESD time difficult}), it is extremely difficult if not impossible to obtain analytic solution of CD time for general reservoir interactions. Here we consider the inverse problem and express the CD
tolerable condition $t_{CD}\geq \tau $ in terms of the double excitation probability, i.e.,
\begin{equation}
D\leq D_{\Phi }^{t}=\frac{\alpha -\beta -\sqrt{(\alpha +\beta )^{2}-4\eta }}{2[f_{2}(\tau )-f_{3}(\tau )]},  \label{ESD-tolerable
Phi-polarization}
\end{equation}
where $\alpha =\rho _{22}f_{4}(\tau )+\rho _{33}f_{1}(\tau )+f_{2}(\tau
)(1-y_{23})$, $\beta =\rho _{22}f_{1}(\tau )+\rho _{33}f_{4}(\tau
)+f_{3}(\tau )(1-y_{23})$, and $\eta =[Q_{\Phi }/2+x_{23}]^{2}|f_{0}(\tau
)|^{2}$. One notes that $D_{\Phi }^{t}$ is controlled by the initial
system parameters: $\rho _{22}$, $\rho _{33}$, and $Q_{\Phi }$ for a given
tolerable time scale $\tau $.

We now make the assumption $p_{k}(t)=1-e^{-\gamma _{k}t}$ with $k=a, b$, and
take $\gamma _{a}=5\gamma _{b}=\gamma $ for convenience. The CD-tolerable critical value $D_{\Phi }^{t}$ is then illustrated in Fig.~\ref{Phi-PhaseDiagram-polarization} by the dashed lines (a) and (b) for $\tau$ equals $1/5\gamma $ and $11/36\gamma $ respectively. Taking $\tau=1/5\gamma $ as an example, the yellow zone (regions \textit{i}, \textit{ii}, \textit{iii}, \textit{iv}) is the CD-tolerable phase with $t_{CD}\geq \tau $, and the red zone (region \textit{v}) is the CD-no-go
phase with $t_{CD}<\tau $. Again, the yellow CD-tolerable region will
decrease with the increase of $\gamma _{a}$, $\gamma _{b}$, $\tau $, see for example a lager value
of $\tau =11/36\gamma $ indicated by line (b). The crossings of the
CD-tolerable critical line with the maximum and minimum physical boundaries
can be obtained by solving the equations $D_{\Phi }^{t}=D_{\Phi }^{\max }$ and $D_{\Phi }^{t}=D_{\Phi }^{\min }$.

We now consider the threshold disentanglement process, which requires $Q_{\Phi }(\tau )\geq C_{tv}$. It can be transformed into a restriction of
the double excitation probability
\begin{equation}
D\leq D_{\Phi }^{tv}=\frac{\alpha -\beta -\sqrt{(\alpha +\beta )^{2}-4\eta
^{\prime }}}{2[f_{2}(\tau )-f_{3}(\tau )]},
\label{TD-tolerablePhi-polarization}
\end{equation}
where $\eta ^{\prime }=[(Q_{\Phi }/2+x_{23})f_{0}(\tau )-C_{tv}/2]^{2}$ with
a $C_{tv}/2$ shift from the CD-tolerable critical value $D_{\Phi }^{t}$.
One notes that $D_{\Phi }^{tv}$ is also controlled by the initial
system parameters $\rho _{22}$, $\rho _{33}$, and $Q_{\Phi }$.

The behavior of $D_{\Phi }^{tv}$ is illustrated by the dot-dashed line in
Fig.~\ref{Phi-PhaseDiagram-polarization} for $\tau =1/5\gamma $ and $C_{tv}=0.1$. Regions \textit{ii}, \textit{iii} is the TD-tolerable phase with $t_{tv}\geq \tau $ and regions \textit{i}, \textit{iv}, \textit{v} is the TD-no-go phase with $t_{tv}<\tau $.

The first type of intuitive and counter-intuitive disentanglement situations can be identified easily by comparing different regions separated by the crossing of a vertical line with various disentanglement critical lines (similar to Fig.~\ref{counter-intuitive}). Fig.~\ref{Phi-PhaseDiagram-polarization} shows explicitly the second type. The counter-intuitive situation is identified by comparing regions \textit{i} and \textit{ii} that are separated by the crossing of the TD-tolerable critical line $D_{\Phi }^{tv}$ and CD-tolerable critical line (b). One notes that although region \textit{ii} suffer earlier CD than region \textit{i}, it reaches TD later. Region \textit{iii} is the intuitive optimal robust phase against both CD and TD.

Thus we have shown, for $\Phi $ type of X-form initial state under
depolarization channel, that there is no CD-free region, and the
CD-tolerable, CD-no-go, TD-tolerable, TD-no-go, and the optimal robust phases are pre-manageable
through the initial system parameters $D$, $\rho _{22}$, $\rho _{33}$, and $Q_{\Phi }$ when the tolerable time scale $\tau $ is given.

\subsubsection{$\Psi $ type of X-form initial state}

\begin{figure}[t]
\includegraphics[width=6cm]{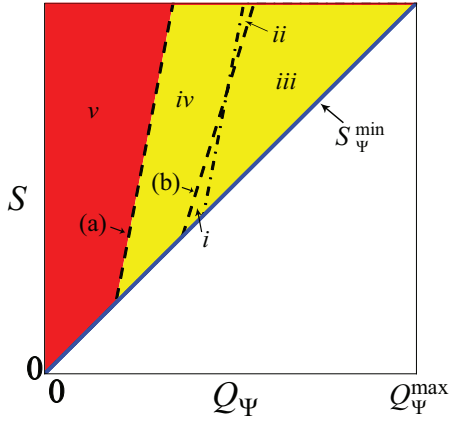}
\caption{Disentanglement phase diagram with respect to single
excitation probability $S$ and entanglement $Q_{\Psi }$ ($C_{0}$) for $\Psi$
type of X-form initial state under
depolarization channel. The CD-tolerable critical value $S_{\Psi }^{t}$ is
illustrated by the dashed lines (a) and (b) for $\tau $ takes $1/9\gamma$ and $1/6\gamma $ respectively. The TD-tolerable
critical value $S_{\Psi }^{tv}$ is illustrated by the dot-dashed line for $\tau =1/9\gamma $ and $C_{tv}=0.1$. For $1/9\gamma$, the yellow zone (regions \textit{i}, \textit{ii}, \textit{iii} and \textit{iv}) and red zone (region \textit{v}) are CD-tolerable (caution) and CD-no-go (danger) phases respectively. Regions \textit{ii} and \textit{iii} form the
TD-tolerable phase, regions \textit{i}, \textit{iv }, \textit{v} form the TD-no-go phase, and region \textit{iii} is
the optimal robust phase.}
\label{Psi-PhaseDiagram-polarization}
\end{figure}

For $\Psi $ type of X-form matrix one has $C_{0}=Q_{\Psi }$. The matrix
positivity requires $Q_{\Psi }/2\leq x_{23}-x_{14}$. Again we consider the physical restriction in terms of the single excitation probability $S=\rho _{22}=\rho _{33}$, and have $S_{\Psi }^{\min }\leq S\leq 1/2$, where the minimum boundary $S_{\Psi }^{\min }$ is the same as given in (\ref{Psi physical-amplitude}). Its behavior is illustrated again in Fig.~\ref{Psi-PhaseDiagram-polarization} by the blue solid line as a function of $Q_{\Psi }$.

From (\ref{CPsi-depolarization}), the CD onset condition is obtained as $|\rho
_{23}(t)|=\sqrt{\rho _{11}(t)\rho _{44}(t)}$. Again, it is extremely difficult to obtain analytic solutions of CD time from this equation. So we consider the inverse problem, and transform the CD tolerable condition $t_{CD}\geq \tau$ in terms of the
single excitation probability bounded by its CD-tolerable critical value $S_{\Psi }^{t}$, i.e.,
\begin{equation}
S\leq S_{\Psi }^{t}=\frac{-\mu +\sqrt{\mu ^{2}-4\lambda \nu }}{2\lambda },
\label{ESD-tolerablePsi-polarization}
\end{equation}
where we have set $\lambda =[f_{2}(\tau )+f_{3}(\tau )]^{2}$, $\mu
=y_{14}[f_{1}(\tau )+f_{4}(\tau )][f_{3}(\tau )+f_{2}(\tau )]$, and $\nu
=\xi -\chi $ with $\xi =[\rho _{11}f_{1}(\tau )+\rho _{44}f_{4}(\tau )][\rho
_{11}f_{4}(\tau )+\rho _{44}f_{1}(\tau )]$,\ $\chi =[Q_{\Psi
}/2+x_{14}]^{2}|f_{0}(\tau )|^{2}$. One sees that $S_{\Psi }^{t}$ is
controlled by the system initial parameters: $\rho _{11}$, $\rho _{44}$, and
$Q_{\Psi }$ for a given tolerable time scale $\tau $.

For demonstration, we make the assumption $p_{k}(t)=1-e^{-\gamma _{k}t}$ with $k=a,b$, and take for convenience $\gamma _{a}=\gamma _{b}=\gamma $ as an illustration. Then the CD-tolerable critical value $S_{\Psi }^{t}$ is illustrated in Fig.~\ref{Psi-PhaseDiagram-polarization} by the dashed lines (a) and (b) for $\tau$ equals $1/9\gamma $ and $1/6\gamma $ respectively. Taking $\tau=1/9\gamma $ as an example, the yellow zone (regions \textit{i}, \textit{ii}, \textit{iii}, \textit{iv}) is the CD-tolerable phase with $t_{CD}\geq \tau $, and the red zone (region \textit{v}) is the CD-no-go phase with $t_{CD}<\tau $. Again, the yellow CD-tolerable region will
decrease with the increase of $\gamma _{a}$, $\gamma _{b}$, $\tau $, see for example a lager value of $\tau =1/6\gamma$ indicated by the dashed line (b). The crossings of the CD-tolerable critical line with the maximum and minimum physical boundaries
can be achieved by solving the equations $S_{\Psi }^{t}=S_{\Psi }^{\max }$ and $S_{\Psi }^{t}=S_{\Psi }^{\min}$.

The threshold disentanglement process requires $Q_{\Psi }(\tau )\geq C_{tv}$. It can be transformed into a restriction of
the single excitation probability
\begin{equation}
S\leq S_{\Psi }^{tv}=\frac{-\mu +\sqrt{\mu ^{2}-4\lambda \nu ^{\prime }}}{2\lambda }
\end{equation}
where $\nu ^{\prime }=\xi -\chi ^{\prime }$ and $\chi ^{\prime }=[(Q_{\Psi
}/2+x_{14})f_{0}(\tau )-C_{tv}/2]^{2}$ with a $C_{tv}/2$ shift from $S_{\Psi
}^{t}$. Obviously it is also determined by the initial parameters $\rho
_{11}$, $\rho _{44}$, and $Q_{\Psi }$.

The behavior of $S_{\Psi }^{tv}$ is illustrated by the dot-dashed line in
Fig.~\ref{Psi-PhaseDiagram-polarization} for $\tau =1/9\gamma $ and $C_{tv}=0.1$. The zone (regions \textit{ii}, \textit{iii}) is the TD-tolerable phase with $t_{tv}\geq \tau $ and the zone (regions \textit{i},
\textit{iv}, \textit{v}) is the TD-no-go phase with $t_{tv}<\tau $.

Again, first type of intuitive and counter-intuitive disentanglement situations can be identified easily by comparing different regions separated by the crossing of a vertical line with various disentanglement critical lines (similar to Fig.~\ref{counter-intuitive}). Fig.~\ref{Psi-PhaseDiagram-polarization} shows explicitly the second type. \textit{i} and \textit{ii} form a pair of counter-intuitive regions, where states with earlier (later) CD counter-intuitively reach TD later (earlier). Region \textit{iii} is the intuitive optimal robust phase against both CD and TD.

Now we can conclude that for $\Psi$ type of X-form initial state under
depolarization channel, there is no CD-free region, and the CD-tolerable,
CD-no-go, TD-tolerable, TD-no-go phases are manageable through the
initial system parameters $S$, $\rho _{11}$, $\rho _{44}$, and $Q_{\Psi }$
when the tolerable time scale $\tau $ is given.

To this end, we have shown explicitly for X-form initial state $\rho_{X}(0)$, its various CD and TD phases are manageable through several
initial parameters under amplitude damping, phase damping, and
depolarization channels. In the next section we will extend this result further to generic two-qubit initial states.

\section{Disentanglement Phases for Generic Mixed States}

We now analyze the CD and TD properties of the generic initial state (\ref{general
initial}) under all three channels. First we discuss some evolution properties of the state. The generic time dependent state is obtained in (\ref{generic time Kraus}), and it can always be expressed as
\begin{equation}
\rho (t)=\left(
\begin{array}{cccc}
\rho _{11}(t) & \rho _{12}(t) & \rho _{13}(t) & \rho _{14}(t) \\
\rho _{21}(t) & \rho _{22}(t) & \rho _{23}(t) & \rho _{24}(t) \\
\rho _{31}(t) & \rho _{32}(t) & \rho _{33}(t) & \rho _{34}(t) \\
\rho _{41}(t) & \rho _{42}(t) & \rho _{43}(t) & \rho _{44}(t)
\end{array}
\right).  \label{generic time}
\end{equation}
Obviously, such a time dependent state can always be decomposed into a sum of
an X-form matrix $X(t)$ and an O-form matrix $O(t)$, i.e.,
\begin{equation}
X(t)=\left(
\begin{array}{cccc}
\rho _{11}(t) & 0 & 0 & \rho _{14}(t) \\
0 & \rho _{22}(t) & \rho _{23}(t) & 0 \\
0 & \rho _{32}(t) & \rho _{33}(t) & 0 \\
\rho _{41}(t) & 0 & 0 & \rho _{44}(t)
\end{array}
\right), \label{X-form time}
\end{equation}
and
\begin{equation}
O(t)=\left(
\begin{array}{cccc}
0 & \rho _{12}(t) & \rho _{13}(t) & 0 \\
\rho _{21}(t) & 0 & 0 & \rho _{24}(t) \\
\rho _{31}(t) & 0 & 0 & \rho _{34}(t) \\
0 & \rho _{42}(t) & \rho _{43}(t) & 0
\end{array}
\right). \label{O-form time}
\end{equation}

Note that in general the evolutions of $X(0)$ and $O(0)$ may have cross dependence, e.g., $X(t)$ may depend on the non-zero initial matrix $O(0)$. Therefore $X(t)$ may not necessarily be identical to the time evolution state $\rho_{X}(t)$ analyzed in previous sections.

\subsection{Independent evolutions of $X(0)$ and $O(0)$ }

It is well known (see examples in Refs.~\cite{Yu-Eberly04, Yu-Eberly07, Bellomo08, Man-etal08, Cui-etal09, Pang-etal12, Duan-etal13}) and is also explicitly shown in Section II, that the evolution of an X-form initial state $\rho _{X}(0)$ retains the X form under all three channels. Here we check whether this property is preserved for the X-form component $X(0)$ of a generic state (\ref{general initial}).

For amplitude damping channel, the corresponding Kraus operators in Eq.~(\ref{Kraus-Amplitude}) are used to obtain the generic time dependent state (\ref{generic time}). It is easy to show that the X-form time dependent component (\ref{X-form time}) is exactly the same as what is given in (\ref{X-amplitude time}), which depends only on the initial X-form matrix $X(0)$. In addition, the matrix elements of the time dependent O-form component (\ref{O-form time}) are given as $\rho _{12}(t)=\rho _{21}^{*}(t)=q_{a}\sqrt{q_{b}}\rho _{12}$, $\rho _{13}(t)=\rho _{31}^{*}(t)=q_{b}\sqrt{q_{a}}\rho _{13}$, $\rho _{24}(t)=\rho _{42}^{*}(t)=\sqrt{q_{a}}\rho _{24}+p_{b}\sqrt{q_{a}}\rho _{13}$, and $\rho _{34}(t)=\rho _{43}^{*}(t)=\sqrt{q_{b}}\rho _{34}+p_{a}\sqrt{q_{b}}\rho _{12}$, which depend only on the initial
O-form component $O(0)$.

Similarly for phase damping channel, it is calculated that $X(t)$ is exactly the same as
in (\ref{X-phase time}), and the O-form matrix elements in (\ref{O-form time}) are given as $\rho
_{12}(t)=\rho _{21}^{*}(t)=\sqrt{q_{b}} \rho _{12}$, $\rho _{13}(t)=\rho
_{31}^{*}(t)=\sqrt{q_{a}}\rho _{13}$, $\rho _{24}(t)=\rho _{42}^{*}(t)=\sqrt{q_{a}}\rho _{24}$, $\rho _{34}(t)=\rho _{43}^{*}(t)=\sqrt{q_{b}}\rho _{34}$. Obviously, the evolutions of $X(0)$ and $O(0)$ have no cross dependence.

Finally for the depolarization channel, the time dependent X-form matrix (\ref{X-form time}) is found to be exactly the same as
Eq.~(\ref{X-polarization time}), which has no dependence on $O(0)$. Also the elements of the time dependent O-form matrix (\ref{X-form time}) have no dependence on $X(0)$, and are obtained as
\begin{equation}
\left(
\begin{array}{c}
\rho _{12}(t) \\
\rho _{13}(t) \\
\rho _{24}(t) \\
\rho _{34}(t)
\end{array}
\right) =\left(
\begin{array}{cccc}
f_{5}(t) & 0 & 0 & f_{6}(t) \\
0 & f_{7}(t) & f_{8}(t) & 0 \\
0 & f_{8}(t) & f_{7}(t) & 0 \\
f_{6}(t) & 0 & 0 & f_{5}(t)
\end{array}
\right) \left(
\begin{array}{c}
\rho _{12} \\
\rho _{13} \\
\rho _{24} \\
\rho _{34}
\end{array}
\right),
\end{equation}
where we have defined $f_{5}(t)=q_{a}q_{b}-q_{a}p_{b}/3+p_{a}q_{b}/3-p_{a}p_{b}/9$, $f_{6}(t)=2p_{a}q_{b}/3-2p_{a}p_{b}/9$, $f_{7}(t)=q_{a}q_{b}+q_{a}p_{b}/3-p_{a}q_{b}/3-p_{a}p_{b}/9$, and $f_{8}(t)=2q_{a}p_{b}/3-2p_{a}p_{b}/9$.

Thus we have shown, for a generic initial state (\ref{general initial}) under all three channels, that the evolutions of components $X(0)$ and $O(0)$ are indeed independent of each other, i.e., for the time dependent components (\ref{X-form time}) and (\ref{O-form time}) one has $X(t)=\sum_{i=1}^{4}K_{i}(t)X(0)K_{i}^{\dag }(t)$ and $O(t)=\sum_{i=1}^{4}K_{i}(t)O(0)K_{i}^{\dag }(t)$ respectively. This property plays a key role in our subsequent discussion of entanglement lower bounds.

\subsection{Disentanglement phases and pre-management}

According to Refs.~\cite{Ma-etal11,Rafsanjani-Agarwal12}, the concurrence
for the static initial generic state (\ref{general initial}) is no less than
that of its X-form component (\ref{X-form initial}), i.e., $C[\rho (0)]\geq
C[X(0)]$. Since we have shown in the previous subsection that all
three quantum channels allow independent evolutions of the X-form and O-form
initial components, the static lower bound relation can then be extended straightforwardly to the time dependent state $\rho (t)=X(t)+O(t)$, i.e.,
\begin{equation}
C[\rho (t)]\geq C[X(t)],
\end{equation}
This means the time dependent concurrence of $\rho (t)$ is always no less than
that of its X-form component at any time $t$. Consequently, the arbitrary
state $\rho (0)$ will always reach CD and TD no
earlier than its X-form component $X(0)$, i.e., $\rho _{X}(0)$.

\begin{figure}[t]
\includegraphics[width=6cm]{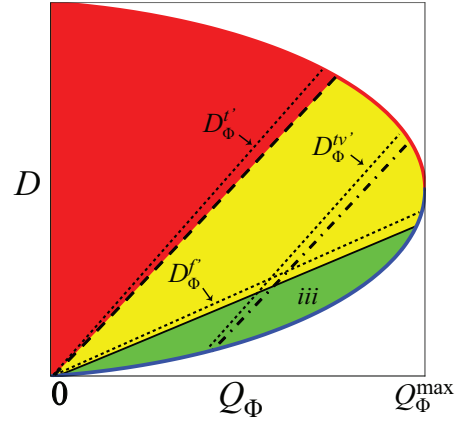}
\caption{Schematic illustration of disentanglement phases for the generic
initial state (\ref{general initial}) under amplitude damping
channel. The CD-free ($D_{\Phi }^{f^{\prime }}$),
CD-tolerable ($D_{\Phi }^{t^{\prime }}$) and TD-tolerable ($D_{\Phi
}^{tv^{\prime }}$) critical values are schematically illustrated by the dotted lines, which are lower bounded by the corresponding
critical lines for the X-form initial state.}
\label{Phi-PhaseDiagram-amplitude-generic}
\end{figure}

This implies that the exact CD-free ($D_{\Phi }^{f}$),
CD-tolerable ($D_{\Phi }^{t}$), and TD-tolerable ($D_{\Phi }^{tv}$)
critical conditions for the X-form state $\rho _{X}(0)$ simply provide lower
bounds for the corresponding exact critical values $D_{\Phi }^{f^{\prime }}$, $D_{\Phi }^{t^{\prime }}$, $D_{\Phi}^{tv^{\prime }}$ of the entire state $\rho (0)$. See Fig.~\ref{Phi-PhaseDiagram-amplitude-generic} for a schematic illustration by the dotted lines. Here we have taken $\Phi $ type of generic state $\rho (0)$ under amplitude damping channel as an example. Various CD and TD phases of the generic state $\rho (0)$ are determined by the dotted critical lines. One notes from Fig.~\ref{Phi-PhaseDiagram-amplitude-generic} that the
CD-free, CD-tolerable, TD-tolerable, as well as the optimal robust (region
\textit{iii}) phases for $\rho (0)$ are lower bounded by the corresponding
phases of $\rho _{X}(0)$. Similar analysis can be applied to $\Psi$ type of $\rho (0)$, as well as other type
of decay channels.

The key information here is that these exact critical
conditions for the X-form state $\rho _{X}(0)$ are sufficient
conditions to guarantee the arbitrary entire state (\ref{general initial})
to be CD free, CD tolerable, TD tolerable, and optimal robust respectively. In this sense,
the pre-management of disentanglement properties of the complex generic state $\rho (0)$ under all three quantum
channels can be simply realized by managing just a few parameters that explicitly determine the corresponding
disentanglement phases of its X-form component $\rho _{X}(0)$.

\section{Summary and Discussion}

In summary we have investigated a new type of disentanglement prevention method, i.e., pre-management with limited partial resources of initial conditions, for a generic two-qubit system that undergoes three typical decoherence channels. The process of threshold disentanglement (TD) is analyzed for the first time, and its intuitive and counter-intuitive connections to initial entanglement and to complete disentanglement (CD) phenomenon are analyzed explicitly. Our analysis illustrates systematically the fundamental connections between open system initial conditions and its disentanglement properties. At the same time, it may also serve as guidance to experimental preparation of robust entangled states.

The disentanglement properties of a general X-form initial state are demonstrated with various exact CD and TD phases in terms of just a few initial parameters, i.e., concurrence and excitation probabilities. These disentanglement phases are shown to have provided lower bounds for the corresponding phases of a generic two-qubit initial state. In this way, the initial parameters of the X-form component amount to a set of management tools that can tune the generic and complex state into particular phases with preferred disentanglement properties. The nature of phase analysis allows a well defined amount of margin for errors in practical imprecise state preparations.

Our treatment of solving initial conditions for a prescribed time scale $\tau$ has avoided much of the difficulty in determining the timing of CD and TD, and thus is capable to deal with many complex situations (e.g., generic two-qubit decay under non-symmetric amplitude damping and depolarization channels) that couldn't be solved analytically in previous studies. Such analysis is well grounded by practical situations when a desired disentanglement time scale is pre-determined by quantum information tasks.

\section{Acknowledgement}

The author would like to acknowledge discussions with J.H. Eberly as well as
partial financial support from the following agencies: DARPA
HR0011-09-1-0008, ARO W911NF-09-1-0385, and NSF PHY-1203931.


\begin{thebibliography}{99}

\bibitem{Yu-Eberly09} T. Yu and J.H. Eberly, \sci{323}, 598 (2009).

\bibitem{Almeida-etal07} M.P. Almeida, F. de Melo, M. Hor-Meyll, A. Salles, S.P. Walborn, P.H. Ribeiro, L. Davidovich, \sci {316}, 579
(2007).


\bibitem{Rau-etal08} A.R.P. Rau, M. Ali, and G. Alber, \epl{82}, 40002 (2008); M. Ali, G. Alber, and A.R.P. Rau, \jpb{42}, 025501 (2009).

\bibitem{ESD-QEC} I. Sainz and G. Bj\"{o}rk , \pra{77}, 052307 (2008); M. Y\"{o}na\c{c} and J.H. Eberly, \qic{14}, 0039 (2014).

\bibitem{Li-etal11} Y. Li, B. Luo, and H. Guo, \pra{84}, 012316 (2011).

\bibitem{Goyal-etal12} S.K. Goyal, S. Banerjee, and S. Ghosh, \pra{85}, 012327 (2012).


\bibitem{Yu-etal} X. Zhao, S.R. Hedemann, T. Yu, arXiv:1305.4627
(2013).

\bibitem{preservation} B. Bellomo, R. Lo Franco, S. Maniscalco, and G. Compagno, \pra{78},
060302(R) (2008); R. Lo Franco, B. Bellomo, E. Andersson, and G. Compagno, \pra{85}, 032318 (2012); B. Bellomo, R. Lo Franco, E. Andersson, J. D. Cresser, and G. Compagno, Phys. Scripta {\bf T147}, 014004 (2012).

\bibitem{preservation-exp} J.-S. Xu, K. Sun, C.-F. Li, X.-Y. Xu, G.-C. Guo, E. Andersson, R. Lo
Franco, and G. Compagno, Nat. Commun. {\bf 4}, 2851 (2013).



\bibitem{Palma96} G.M. Palma, K.A. Suominen, and A.K. Ekert, Proc.
R. Soc. London, Ser. A \textbf{452}, 567 (1996).

\bibitem{Zanardi-Rasetti97} P. Zanardi and M.
Rasetti, \prl{79}, 3306 (1997).

\bibitem{Lidar98} D.A. Lidar, I.L. Chuang, and K.B. Whaley, \prl{81}, 2594 (1998).

\bibitem{Yu-Eberly04} T. Yu and J.H. Eberly, \prl{93}, 140404 (2004), and  T. Yu and J.H. Eberly, \prl{97}, 140403 (2006).

\bibitem{Ann-Jaeger07} K. Ann and G. Jaeger, \pra{76}, 044101 (2007).

\bibitem{Lopez08} C.E. L\'{o}pez, G. Romero, F. Lastra, E. Solano, and J.C. Retamal, \prl{101}, 080503 (2008).

\bibitem{Novotny11} J. Novotn\'{y}, G. Alber, and I. Jex, \prl{107} 090501 (2011).

\bibitem{Julian-etal12} K.M. Fonseca-Romero, J. Mart\'{i}nez-Rinc\'{o}n, and C. Viviescas, \pra{86}, 042325 (2012).

\bibitem{Franco-etal13} R. Lo Franco, B. Bellomo, S. Maniscalco, and G. Compagno, Int. J. Mod. Phys. B {\bf 27}, 1345053 (2013).

\bibitem{Yu-Eberly07} T. Yu and J.H. Eberly, \qic{7}, 459 (2007).

\bibitem{Bellomo08} B. Bellomo, R. Lo Franco, and G. Compagno, \pra{77}, 032342 (2008).

\bibitem{Man-etal08} Z.-X. Man, Y-J Xia, and N.B. An, \jpb{41}, 085503 (2008).

\bibitem{Cui-etal09} W. Cui, Z. Xi, and Y. Pan, \jpa{42}, 025303 (2009).

\bibitem{Pang-etal12} C.-Q. Pang, F.-L. Zhang, Y. Jiang, M.-L. Liang, J.-L. Chen, \qic{13}, 0645 (2013).

\bibitem{Duan-etal13} L. Duan, H, Wang, Q.-H. Chen, and Y. Zhao, J. Chem. Phys. {\bf 139}, 044115 (2013).

\bibitem{Huang-Zhu07} J.-H. Huang and S.-Y. Zhu, \pra{76}, 062322 (2007); J.-H. Huang and S.-Y. Zhu, \oc {281}, 2156 (2008).

\bibitem{Fanchini-etal11} F.F. Fanchini, P.E.M.F. Mendon\c{c}a, R.d.J. Napolitano, \qic{11}, 0677 (2011).

\bibitem{Qian-Eberly12} X.-F. Qian and J.H. Eberly, \pla{376}, 2931 (2012); Xiao-Feng Qian, Conference paper W6.25, CQO-X and QIM-2 \copyright OSA (2013).

\bibitem{Yang-etal13} B.-Y. Yang, M.-F. Fang , and J. Huang, Chin. Phys. B {\bf 22}, 080303 (2013).


\bibitem{Schmidt} R. Grobe, K. Rz\c{a}zewski and J.H. Eberly, \jpb {27}, L503 (1994). See also A. Ekert and P.L. Knight, \ajp{63}, 415
(1995) and J.H. Eberly, Laser Phys. {\bf 16}, 921 (2006).

\bibitem{Wootters} W.K. Wootters, \prl{80}, 2245
(1998).

\bibitem{negativity} G. Vidal and R.F. Werner, \pra{65}, 032314
(2002).

\bibitem{Ma-etal11} Z.-H. Ma, Z.-H. Chen, J.-L. Chen, C. Spengler, A.
Gabriel, and M. Huber, \pra{83}, 062325 (2011).

\bibitem{Rafsanjani-Agarwal12} S.M. Hashemi Rafsanjani and S. Agarwal, arXiv: 1204.3912 (2012); S.M. Hashemi Rafsanjani, M. Huber, C.J. Broadbent, and J. H. Eberly, \pra{86}, 062303 (2012).

\bibitem{Kimble-etal09} S.B. Papp, K.S. Choi, H. Deng, P. Lougovski, S.J. van Enk, H. J.
Kimble, \sci{324}, 764 (2009).

\bibitem{Davidovich-etal12} O.J. Far\'{i}as, G.H. Aguilar, A. Vald\'{e}s-Hern\'{a}ndez, P.H. Ribeiro, L. Davidovich, and S.P. Walborn, \prl{109}, 150403 (2012).

\bibitem{Huber13} M. Huber and J.I. de Vicente, \prl{110}, 030501 (2013).


\bibitem{NC-Preskill} M.A. Nielsen and I.L. Chuang, \emph{Quantum
Computation and Quantum Information} (Cambridge Univ. Press, 2000); and J.
Preskill, \emph{Quantum Information and Computation}, Caltech Lecture Notes
for Ph219/CS219.

\bibitem{Horodecki09-etal} R. Horodecki, P. Horodecki, M. Horodecki, and K. Horodecki, \rmp{81}, 865 (2009).


\end{thebibliography}
\end{document}